\documentclass[11pt]{article}
\usepackage[preprint]{acl}
\usepackage{times}
\usepackage{latexsym}
\usepackage[T1]{fontenc}
\usepackage[utf8]{inputenc}
\usepackage{microtype}
\usepackage{inconsolata}
\usepackage{graphicx}
\usepackage{amsmath,amssymb}
\usepackage{booktabs}
\usepackage{colortbl}
\usepackage{multirow}
\usepackage{tabularx}
\graphicspath{{figures/}}

\usepackage{xcolor}

\definecolor{tableOneBest}{HTML}{E1F0D6}
\definecolor{tableOneSecond}{HTML}{FFF5CF}
\usepackage{placeins}

\usepackage{needspace}

\title{DuoLM: Fast and Effective Personalized Recommendation with Language Models}
\author{
  Jiang Wu\textsuperscript{1}\thanks{Equal contribution.}\quad
  Yutong Song\textsuperscript{2}\footnotemark[1]\quad
  Hongbo Wang\textsuperscript{3}\quad
  Guanghe Cao\textsuperscript{3}\quad
  Jeff Lei Liang\textsuperscript{4}\\
  \textsuperscript{1}Independent Researcher\quad
  \textsuperscript{2}University of California, Irvine\\
  \textsuperscript{3}Google DeepMind\quad
  \textsuperscript{4}Oracle
}

\newif\ifreferencesready
\referencesreadytrue

\begin{document}
\maketitle

\begin{abstract}
Language models can make personalized recommendations from rich item content, user history, and feedback. However, evaluating many candidates with full-text inputs incurs substantial online computation and limits serving throughput. We introduce DuoLM, a framework for fast and effective personalized recommendation that integrates a compact recommendation model with an efficient serving runtime. An encoder LLM learns compact item representations for offline reuse, while a personalized ranker LLM combines them with user history and feedback for recommendation. We train both models end to end with recommendation supervision and teacher distillation. The runtime further accelerates inference through virtual token IDs for embedding access, shared history computation, and CUDA Graph replay. Benchmark results show that our framework achieves recommendation quality comparable to that of the expert teacher model. On a single H100, it delivers approximately 26,400 candidate scores per second, about $60\times$ the throughput of the teacher served with vLLM and automatic prefix caching.

\end{abstract}

\section{Introduction}
\label{sec:introduction}

\begingroup\setlength{\fboxsep}{7pt}\setbox0=\hbox{\colorbox{white}{\parbox{\dimexpr\linewidth-2pt-2\fboxsep\relax}{\small\raggedright
\small\itshape
``Nature is an endless combination and repetition of a very few laws. She hums the old well-known air through innumerable variations.''
\upshape
\mbox{-- Ralph Waldo Emerson, \textit{History}}
}}}\noindent{\color[rgb]{0.290,0.325,0.337}\vrule width 2pt height \ht0 depth \dp0}\usebox0\par\endgroup\smallskip

From personalized shopping to content discovery, language models use item content, user histories, and feedback to assess which items best match individual preferences~\citep{bao2023tallrec,geng2022p5,yue2023llamarec}. Each recommendation request typically involves multiple candidates, while the system must sustain a continuous stream of requests within a given computational budget. The practical value of language models for recommendation therefore depends on both recommendation quality and serving efficiency.

Despite advances in language-model-based personalized recommendation, efficient online serving remains challenging. We identify two primary challenges.

\textit{First, it remains difficult to substantially reduce online computation without compromising recommendation quality.} Traditional ID-based two-tower recommenders score items efficiently, but models trained solely on interactions do not directly incorporate textual semantics or pretrained world knowledge. This reliance on interaction data limits generalization to long-tail and cold-start items~\citep{hou2022universalsequencerepresentationlearning,li2023textneedlearninglanguage,kim2024allmrec}. Full-text cross-encoders use item descriptions together with user history and feedback, but the resulting candidate representations depend on the user context and cannot be directly reused across users. Compressing item text can reduce online computation~\citep{ge2024incontextautoencodercontextcompression,chevalier2023autocompressors}, yet may discard details that matter to particular users. The challenge is to obtain compact, reusable item representations that still support high-quality personalized judgments.

\textit{Second, a reduction in model computation does not automatically translate into faster end-to-end request processing.} Inference engines such as vLLM~\citep{kwon2023efficient} improve latency and throughput via batching and prefix caching, but redundant work can remain in how a recommendation service prepares and submits its workload. A single recommendation request typically evaluates multiple candidates against the same user history. If candidates are submitted separately, the service may repeatedly construct and transmit the same history inputs and incur separate scheduling overhead for each candidate. Prefix caching can reuse history KV states~\citep{juravsky2024hydragen}, but does not per se eliminate these repeated operations. As model execution becomes faster, these overheads can limit end-to-end recommendation throughput.

\begin{figure*}[!t]
\centering
\includegraphics[width=0.9\textwidth]{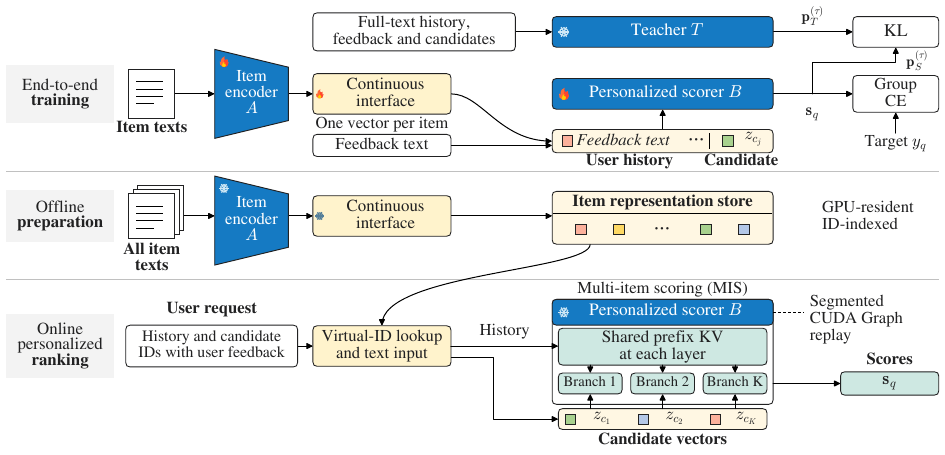}
\vspace{-0.5em}
\caption{\textbf{Overview of DuoLM.} Encoder $A$ and ranker $B$ are trained end to end through a continuous interface, using recommendation labels and distillation from a frozen teacher $T$. After training, we precompute one reusable vector per item. Online, virtual IDs retrieve these vectors for $B$, which scores candidates using user history and feedback. History KV is shared within each request, while candidate branches remain independent.}
\label{fig:framework-overview}
\end{figure*}

To address these challenges, we introduce DuoLM, a framework that integrates representation learning and efficient inference for personalized recommendation (Figure~\ref{fig:framework-overview}). \textbf{The core idea is to learn reusable item representations and personalization together, enabling offline item encoding. Runtime optimizations translate the resulting computational savings into faster serving.}

\textit{On the model side,} an encoder LLM compresses each item's text into a single hidden-state vector, which replaces its text tokens in a personalized ranker LLM. The ranker receives a candidate vector with individual history-item vectors and their associated textual feedback. Each candidate attends to these history entries to produce personalized score.

We co-train the encoder and ranker end to end with recommendation labels and distillation from a frozen full-text teacher~\citep{hinton2015distillingknowledgeneuralnetwork}. Cross-entropy encourages selection of the observed target, while KL divergence encourages the compact model to match the teacher's preference distribution over the same candidates. After training, we precompute item representations offline and run only the ranker online.

\textit{On the runtime side}, we introduce a series of coordinated optimizations tailored to this model architecture. First, \textbf{virtual token IDs (VIDs)} reference a GPU-resident table of precomputed item embeddings shared by history and candidate items. The runtime resolves these IDs into continuous input embeddings for the ranker. Requests therefore carry item IDs and user feedback instead of full item vectors. In our rating benchmark, VIDs reduce average request payloads by over 90\% compared with full-vector transmission.

We build a novel \textbf{multi-item scoring (MIS)} path on SGLang~\citep{NEURIPS2024_724be447} to score multiple candidates in one request. The runtime constructs and prefills the user history once, then shares its prefix KV cache across candidate branches. A tree-structured attention mask allows each candidate to attend only to the shared history and its own branch, preserving independent scoring.

We further adapt the MIS scoring path to segmented prefill \textbf{CUDA Graphs}. These graphs capture GPU operations and their dependencies. The runtime replays the captured graphs to reduce CPU scheduling and kernel launch overhead.

Experiments on Amazon Books and Video Games show: \textit{(i)} Our model outperforms the expert teacher model on Books recommendation (NDCG@10: 0.5586 vs.\ 0.5440), with comparable performance on Video Games (0.5298 vs.\ 0.5313) and Books rating prediction. \textit{(ii)} On a single H100,  recommendation throughput reaches up to $60.1\times$ and $109.9\times$ that of the teacher served by vLLM and SGLang, respectively. Our contributions can be summarized as follows:\\
1. We introduce DuoLM, a framework that learns compact, reusable, efficient representations for personalized recommendation.\\
2. We develop an efficient serving runtime that reuses item representations across users and history computation across candidates.\\
3. We demonstrate substantial throughput gains with recommendation quality comparable to that of the expert teacher model, and share practical lessons from model and runtime ablations.

\section{Method}
\label{sec:method}

\paragraph{Overview.}
For request $q$, we score $K_q$ candidate items using an ordered history of $n_q$ interactions:
\[
\begin{aligned}
H_q &= ((i_1,r_1),\ldots,(i_{n_q},r_{n_q})),\\
C_q &= (c_1,\ldots,c_{K_q}).
\end{aligned}
\]
Each history entry $(i_t,r_t)$ records an item and the user's feedback on it.
As shown in Figure~\ref{fig:framework-overview}, an encoder LLM $A$ maps item text to user-independent representations.
A continuous input interface converts each representation into one input embedding for a causal ranker LLM $B$, which scores candidates using the ordered history of item embeddings and their associated feedback text (Section~\ref{sec:scoring}).
The encoder, interface, and ranker are jointly trained with recommendation labels and distillation of candidate distributions from a separately trained, frozen full-text teacher $T$ (Section~\ref{sec:learning}).
After training, we precompute item embeddings and run only $B$ online.
Our runtime reuses these embeddings across requests and shares history KV states across candidates within each request (Section~\ref{sec:serving}).

\subsection{Model Architecture}
\label{sec:scoring}

The item encoder $A$ produces user-independent content representations,
while the personalized ranker $B$ scores candidates using these
representations together with user history and feedback.
A continuous input interface connects the two models, both initialized
from pretrained language models with hidden dimension $d$.

\paragraph{Encoder LLM.}
Let $x_i$ denote item $i$'s templated token sequence, terminated by EOS.
We use the final-layer EOS hidden state as the item representation,
$a_i=\big[A_{\theta_A}(x_i)\big]_{\mathrm{last}}\in\mathbb{R}^d$,
where $[\cdot]_{\mathrm{last}}$ selects the last valid input position.

\paragraph{Continuous Input Interface.}
The interface maps $a_i$ to a single continuous embedding $z_i$,
replacing the item's text span in $B$.
History and candidate items share this interface.
We RMS-normalize $a_i$ and calibrate its scale against $B$'s native
token embeddings.

For $v\in\mathbb{R}^d$, define
$\operatorname{RMS}(v)=\sqrt{d^{-1}\sum_{k=1}^{d}v_k^2}$.
The interface computes
\begin{equation}
\widehat a_i
=
\frac{a_i}{\sqrt{\operatorname{RMS}(a_i)^2+\epsilon}},
\quad
z_i=\rho e^\gamma\,\widehat a_i.
\label{eq:content-map}
\end{equation}
We estimate the reference scale $\rho>0$ from $B$'s initial native
embedding table $E_B^{(0)}$:
\begin{equation}
\rho^2
=
\frac{1}{|\mathcal U|}
\sum_{u\in\mathcal U}
\operatorname{RMS}\!\left(E_B^{(0)}[u]\right)^2,
\label{eq:rms-reference}
\end{equation}
where $\mathcal U$ is the multiset of token occurrences in sampled
training-item texts.
We fix $\rho$, initialize the learnable scalar log-gain $\gamma$
to zero, and use $\epsilon>0$ for numerical stability.
\begin{figure*}[!t]
\centering
\includegraphics[width=\textwidth]{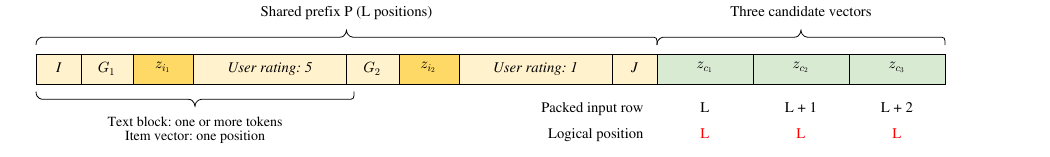}
\vspace{-1em}
\caption{\textbf{Mixed inputs and candidate positions of MIS.} The shared prefix \(P\) contains prompt text, history-item vectors, and ratings. The attention mask allows each candidate to attend to \(P\) and itself, while masking all other candidates. All three use logical position \(L\), the first position after \(P\). }
\label{fig:input-layout}
\end{figure*}
\paragraph{Personalized Ranker LLM.}
The ranker $B$ receives continuous item embeddings and textual feedback.
For each candidate, its input comprises an ordered history prefix
followed by the candidate embedding.

Let $\mathcal E_B(v)$ denote $B$'s embedding sequence for a separately
tokenized text segment $v$, and let $\oplus$ denote concatenation.
We construct the $t$-th history entry from its marker $G_t$,
item embedding $z_{i_t}$, and feedback text $F(r_t)$:
\[
R_{q,t}
=
\mathcal E_B(G_t)
\oplus z_{i_t}
\oplus \mathcal E_B(F(r_t)).
\]
An instruction $I$ and candidate header $J$ delimit the ordered history:
\begin{equation}
\begin{aligned}
P_q
&=
\mathcal E_B(I)
\oplus R_{q,1}\oplus\cdots\oplus R_{q,n_q}\\
&\quad\oplus \mathcal E_B(J).
\end{aligned}
\label{eq:prefix}
\end{equation}
Each feedback segment immediately follows its corresponding item
embedding.
Appending $z_{c_j}$ yields the following mixed input, where boxes
denote continuous embeddings:
\par\smallskip
\begingroup
\small\ttfamily
\renewcommand{\arraystretch}{0.95}
\noindent
\begin{tabular}{@{}l@{\ }c@{\ }l@{}}
\multicolumn{3}{@{}l@{}}{\textless instruction\textgreater}\\
History 1: Item & : & \ensuremath{\boxed{z_{i_1}}}\\
User rating    & : & \textless rating\textgreater\\
\multicolumn{3}{@{}l@{}}{\ensuremath{\ldots}}\\
Candidate: Item & : & \ensuremath{\boxed{z_{c_j}}}
\end{tabular}
\par
\endgroup
\smallskip

Let $L_q$ denote the prefix length, counting each item embedding
as one position.
A shared linear head $w\in\mathbb{R}^d$ scores the candidate's
final-layer hidden state, conditioned on $P_q$ through causal attention:
\begin{equation}
\begin{aligned}
h_{q,j}
&=
\big[B_{\theta_B}(P_q\oplus z_{c_j})\big]_{\mathrm{last}},\\[4pt]
s_{q,j}
&=
w^{\mathsf T}h_{q,j}.
\end{aligned}
\label{eq:score}
\end{equation}
While $z_{c_j}$ represents user-independent item content,
$h_{q,j}$ incorporates the current user's history and feedback.
Each candidate is scored with the shared prefix, without
cross-candidate attention.
The scores $\mathbf s_q=(s_{q,1},\ldots,s_{q,K_q})$ directly rank
candidates without autoregressive decoding.

\subsection{Training}
\label{sec:learning}

For each training request $q$, let $K_q$ denote the number of candidates and $y_q$ the index of the observed target.

\paragraph{Full-text Teacher.}

The teacher $T$ receives the same instruction, ordered history, and feedback as $B$, replacing item vectors with their original texts. Its input ends with a scoring suffix:
\par\smallskip
\begingroup
\small\ttfamily
\renewcommand{\arraystretch}{0.95}
\noindent
\begin{tabular}{@{}l@{\ }c@{\ }l@{}}
\multicolumn{3}{@{}l@{}}{\textless instruction\textgreater}\\
History 1: Item & : & \textless item text\textgreater\\
User rating & : & \textless rating\textgreater\\
\multicolumn{3}{@{}l@{}}{\ensuremath{\ldots}}\\
Candidate: Item & : & \textless candidate text\textgreater\\
Next-book score & : &
\end{tabular}
\par
\endgroup
\smallskip

A linear head maps the final-layer hidden state at the last token of the scoring suffix to a scalar score $t_{q,j}$.
We train $T$ with cross-entropy to identify the observed target within each candidate group.
After training, we freeze $T$ and cache its raw scores for the exact histories and ordered candidate groups used in student training.

\paragraph{End-to-end Training.}
We use these cached scores together with the observed targets to jointly train $A$, the continuous input interface, and $B$. The trainable parameters are $\Theta=(\theta_A,\gamma,\theta_B,w)$, including both backbones' native embeddings; the teacher and reference scale $\rho$ remain fixed. Item vectors are recomputed during each training forward pass, preserving the differentiable path from the scoring model through the interface to the encoder.

To transfer the teacher's relative preferences within each candidate group, we convert teacher scores $t_{q,j}$ and student scores $s_{q,j}$ into probability distributions using a shared temperature $\tau>0$:
\begin{equation}
\begin{aligned}
p_{\mathrm T,q,j}^{(\tau)}
&=\frac{\exp(t_{q,j}/\tau)}
{\sum_{k=1}^{K_q}\exp(t_{q,k}/\tau)},\\[4pt]
p_{\mathrm S,q,j}^{(\tau)}
&=\frac{\exp(s_{q,j}/\tau)}
{\sum_{k=1}^{K_q}\exp(s_{q,k}/\tau)}.
\end{aligned}
\label{eq:distributions}
\end{equation}
Both denominators sum over all $K_q$ candidates in the same request. These distributions define the distillation loss, while the observed target provides direct recommendation supervision:
\begin{equation}
\begin{aligned}
\ell_{\mathrm{CE}}(q)
&=-\log p_{\mathrm S,q,y_q}^{(1)},\\[4pt]
\ell_{\mathrm{KD}}(q)
&=\tau^2\sum_{j=1}^{K_q}p_{\mathrm T,q,j}^{(\tau)}
\log\frac{p_{\mathrm T,q,j}^{(\tau)}}
{p_{\mathrm S,q,j}^{(\tau)}}.
\end{aligned}
\label{eq:student-loss-components}
\end{equation}
The recommendation loss uses temperature one. The distillation loss is the teacher-to-student KL divergence with the standard $\tau^2$ scaling~\citep{hinton2015distillingknowledgeneuralnetwork}. At update $u$, we combine the two losses with distillation weight $\beta_u$:
\begin{equation}
\ell_u(q)
=(1-\beta_u)\ell_{\mathrm{CE}}(q)
+\beta_u\ell_{\mathrm{KD}}(q).
\label{eq:student-objective}
\end{equation}
For a batch $\mathcal B$ of training requests, we minimize $\mathcal L_S(\Theta;u)=|\mathcal B|^{-1}\sum_{q\in\mathcal B}\ell_u(q)$, giving each request equal weight.
Appendix~\ref{app:score-gradients} gives the corresponding score gradients.

The temperature $\tau$ controls the concentration of the distillation distributions, while $\beta_u$ controls the relative contributions of the two supervision terms. We linearly increase the distillation weight $\beta_u$ to
$\beta_{\mathrm{final}}$ and then keep it fixed
(Appendix~\ref{app:distillation-schedule}).


\subsection{Efficient Inference}
\label{sec:serving}

\paragraph{Offline Item Representations and Virtual Token IDs.}
After freezing the student, we export its post-interface vectors to a GPU-resident table $D$. Let $\iota(i)\in\{0,\ldots,N-1\}$ identify item $i$'s unique row and $V$ be the native vocabulary size. We store $D_{\iota(i)}=\operatorname{FP16}(z_i)$ and assign the virtual input ID $V+\iota(i)$ to this row. The embedding lookup returns ${\Lambda(V+\iota(i))}=D_{\iota(i)}$ for virtual IDs and $\Lambda(v)=E_B[v]$ for native IDs $0\le v<V$, using the trained frozen table $E_B$. The same item row serves history and candidate occurrences across requests. Online inputs contain IDs and feedback text, avoiding item encoding and full-vector transmission. Scores still use the scalar head; virtual IDs do not expand the output vocabulary.

\paragraph{Multi-item Scoring with Shared Prefix KV.}
We pack candidates after a shared prefix to reuse history computation, forming $X_q=P_q\oplus z_{c_1}\oplus\cdots\oplus z_{c_{K_q}}$. A standard causal mask permits cross-candidate attention, while consecutive position IDs assign candidates different logical positions. Our multi-item scoring (MIS) path on SGLang~\citep{NEURIPS2024_724be447} controls both visibility and position IDs to preserve per-candidate scoring semantics (Appendix~\ref{app:mis-extensions}).

\begin{figure}[!t]
\centering
\includegraphics[width=0.84\columnwidth]{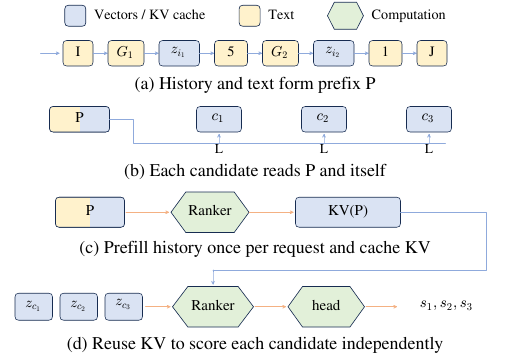}
\vspace{-1.0em}
\caption{\textbf{Multi-item scoring with shared history.}
The same ranker reuses the prefix KV cache to score candidates independently with a shared scoring head.}
\label{fig:scoring-example}
\end{figure}

Let $\mathcal V_q(a)$ denote the key rows visible to query row $a$, and $\pi_q(a)$ its logical position. Prefix rows retain $\pi_q(a)=a$ for $0\le a<L_q$. Candidate $j$ occupies packed row $a_j=L_q+j-1$ but uses $\pi_q(a_j)=L_q$, matching standalone scoring. Visibility is defined by
\begin{equation}
\begin{aligned}
\mathcal V_q(a)&=\{0,\ldots,a\},\qquad 0\le a<L_q,\\
\mathcal V_q(a_j)&=\{0,\ldots,L_q-1\}\cup\{a_j\}.
\end{aligned}
\label{eq:candidate-visibility}
\end{equation}
All other edges, including cross-request edges, are masked. Position-dependent operations use $\pi_q$. These rules describe semantic rows; transport delimiters and graph padding are handled separately. Figure~\ref{fig:input-layout} illustrates the layout; $P,L$ abbreviate $P_q,L_q$ in this figure and Figure~\ref{fig:scoring-example}.

Since the prefix does not attend to candidates, $B$ prefills it once per request and reuses its layerwise KV cache across candidate branches (Figure~\ref{fig:scoring-example}). Item embeddings are reused across requests, whereas prefix KV is shared within each request.

Appendix~\ref{app:mis-attention} gives the corresponding attention formulation.
These visibility and position rules apply at every layer, giving each candidate the same attention context and logical positions as in standalone scoring. Tokenwise normalization and feed-forward layers preserve candidate isolation. A shared linear head scores the final candidate states and returns scores in input order.

Let $s_{q,j}^{\mathrm{MIS}}$ denote the scalar readout at packed row $a_j$, and $s_{q,j}$ the standalone score in Eq.~\eqref{eq:score}. For the same full-attention ranker with identical input embeddings and dropout disabled, these rules imply, in exact arithmetic,
\begin{equation}
s_{q,j}^{\mathrm{MIS}}=s_{q,j},\qquad j=1,\ldots,K_q,
\label{eq:scoring-equivalence}
\end{equation}
including when both layouts use identically rounded item embeddings. 

For the same compact model, standalone scoring without history reuse and MIS have the following numbers of allowed query--key pairs per layer:
\begin{equation}
\begin{aligned}
N_{\mathrm{sep},q}
&=K_q\frac{L_q(L_q+1)}{2}+K_q(L_q+1),\\
N_{\mathrm{MIS},q}
&=\frac{L_q(L_q+1)}{2}+K_q(L_q+1).
\end{aligned}
\label{eq:attention-counts}
\end{equation}
MIS removes $(K_q-1)L_q(L_q+1)/2$ repeated prefix pairs while retaining all candidate--history pairs. These semantic counts exclude delimiters, padding, and kernel overheads.

\paragraph{Enable CUDA Graph.}
After prefix reuse, We further adapt segmented prefill capture to MIS using CUDA Graphs. The runtime replays the captured GPU operations with current inputs and updated candidate position IDs, reducing repeated per-kernel host dispatch. Request-specific handling of prefix and candidate boundaries, remains outside capture.


\section{Experiment Settings}
\label{sec:experimental-setup}

\begin{table*}[t]
\centering
\begingroup
\small
\definecolor{tableOneBest}{HTML}{E1F0D6}
\definecolor{tableOneSecond}{HTML}{FFF5CF}
\setlength{\tabcolsep}{3.6pt}

\caption{Rating prediction on Books. All four history lengths use the same users and targets. Average metrics weight each $K$ equally. Lower is better for all metrics ($\downarrow$). Green and yellow mark the best and second-best results in each column, respectively. MAE$_{\mathrm{med}}$ uses the predictive median; MAE$_{\mathrm{exp}}$ and RMSE use the expected rating.}
\label{exp:rating-quality}

\vspace{-0.5em}
\renewcommand{\arraystretch}{1.0}
\begin{tabular*}{\textwidth}{@{\extracolsep{\fill}}l*{9}{r}@{}}
    \toprule
    & \multicolumn{4}{c}{\textbf{NLL by history length} $\downarrow$}
    & \multicolumn{5}{c}{\textbf{Average across history lengths}} \\
    \cmidrule(lr){2-5}
    \cmidrule(lr){6-10}
    \textbf{Method}
    & $K=1$ & $K=3$ & $K=5$ & $K=10$
    & \textbf{NLL} $\downarrow$
    & \textbf{MAE}$_{\mathrm{med}}$ $\downarrow$
    & \textbf{MAE}$_{\mathrm{exp}}$ $\downarrow$
    & \textbf{RMSE} $\downarrow$
    & \textbf{RPS} $\downarrow$ \\
    \midrule
    Global prior
    & 1.0401 & 1.0401 & 1.0401 & 1.0401
    & 1.0401 & 0.6017 & 0.7824 & 0.8858 & 0.1125 \\
    Item prior
    & 1.0358 & 1.0358 & 1.0358 & 1.0358
    & 1.0358 & 0.6020 & 0.7760 & 0.8813 & 0.1115 \\
    History prior
    & 1.0151 & 0.9706 & 0.9460 & 0.9147
    & 0.9616 & 0.5564 & 0.6919 & 0.8041 & 0.0994 \\
    Content kNN
    & 1.0151 & 0.9707 & 0.9460 & 0.9147
    & 0.9616 & 0.5566 & 0.6919 & 0.8040 & 0.0994 \\
    \addlinespace[3pt]
    Raw Qwen3
    & 1.9020 & 1.3401 & 1.2407 & 1.1103
    & 1.3983 & 0.6581 & 0.7144 & 0.8525 & 0.1257 \\
    HLLM-rating
    & 1.0080 & 0.9543 & 0.9373 & 0.9225
    & 0.9555 & 0.5477 & 0.7110 & 0.8324 & 0.1009 \\
    \addlinespace[3pt]
    DuoLM-FIE (m=8)
    & \cellcolor{tableOneBest}0.9796
    & \cellcolor{tableOneBest}0.9318
    & \cellcolor{tableOneBest}0.9039
    & 0.8790
    & \cellcolor{tableOneSecond}0.9236
    & \cellcolor{tableOneSecond}0.5311
    & \cellcolor{tableOneSecond}0.6434
    & \cellcolor{tableOneSecond}0.7736
    & \cellcolor{tableOneSecond}0.0965 \\
    DuoLM-FIR (m=8)
    & 0.9831 & 0.9331 & 0.9053
    & \cellcolor{tableOneSecond}0.8780
    & 0.9249 & 0.5319 & 0.6472 & 0.7778 & 0.0966 \\
    DuoLM-FIR (m=1)
    & 0.9848 & 0.9337 & 0.9061 & 0.8806
    & 0.9263 & 0.5362
    & \cellcolor{tableOneBest}0.6415
    & \cellcolor{tableOneBest}0.7725
    & 0.0968 \\
    \midrule
    Teacher model
    & \cellcolor{tableOneSecond}0.9810
    & \cellcolor{tableOneSecond}0.9325
    & \cellcolor{tableOneSecond}0.9041
    & \cellcolor{tableOneBest}0.8758
    & \cellcolor{tableOneBest}0.9233
    & \cellcolor{tableOneBest}0.5308
    & 0.6515 & 0.7795
    & \cellcolor{tableOneBest}0.0962 \\
    \bottomrule
\end{tabular*}
\endgroup
\end{table*}

\paragraph{Data and evaluation.}
We evaluate rating prediction on Books and next-item recommendation on Books and Video Games from Amazon Reviews 2023~\citep{hou2024bridging}.
Rating prediction uses 10,000 training users and 1,000 development users, with history lengths $K\in\{1,3,5,10\}$.
We report NLL, MAE$_{\mathrm{med}}$, MAE$_{\mathrm{exp}}$, RMSE, and RPS.
MAE$_{\mathrm{med}}$ uses the predictive median, whereas MAE$_{\mathrm{exp}}$ and RMSE use the expected rating.
For next-item recommendation, each dataset uses 50,000 training users, yielding 237,558 training queries for Books and 233,868 for Video Games.
We report HR@1, HR@10, and NDCG@10 on 5,000 Books users and 3,000 Video Games users.
Data splits, candidate construction, and detailed evaluation protocols appear in Appendix~\ref{app:data-eval}.

\textbf{Model configuration.}
For fair comparisons, we use 0.6B Qwen3 backbones across methods, with the trained full-text teacher as the quality reference for both tasks.
\textbf{Feedback-in-Encoder (FIE)} encodes feedback together with item content, whereas \textbf{Feedback-in-Ranker (FIR)} supplies feedback as text to the Ranker.
Our default, DuoLM ($m=1$), uses FIR with one vector per item and the short input-suffix template.
Model and training details appear in Appendices~\ref{app:models} and~\ref{app:training}, respectively.

\paragraph{Rating baselines.}
i) Global prior uses the overall training-set rating distribution;
ii) Item prior uses the target item's rating distribution among training users;
iii) History prior uses the current user's historical rating distribution;
iv) Content kNN weights historical ratings by the content similarity between each history item and the target;
v) Raw Qwen3-0.6B reads the user history, feedback, and target text to predict a five-star rating;
vi) HLLM-rating adapts HLLM to five-star prediction.

\paragraph{Recommendation baselines.}
i) Popularity ranks items by the number of distinct training users who interacted with them;
ii) History similarity compares each candidate with the average content vector of the user's history;
iii) ContentSASRec, based on SASRec~\citep{kang2018sasrec}, models historical item vectors, ratings, and their order with a sequential Transformer to score candidates.
iv) HLLM~\citep{chen2024hllmenhancingsequentialrecommendations} encodes item content and user history separately, then ranks candidates by user--item representation similarity.
Baseline implementations details in Appendix~\ref{app:models}.
\section{Results and Analysis}
\label{sec:results}

\subsection{Recommendation Quality}
\label{sec:quality}
\label{sec:recommendation-quality}

\paragraph{RQ1: Can our framework maintain high-quality personalized rating prediction and recommendation?}

\paragraph{Rating prediction.}
Table~\ref{exp:rating-quality} shows that DuoLM achieves rating prediction quality close to that of the full-text teacher.
Its average NLL across history lengths is 0.9263, compared with the teacher's 0.9233.
DuoLM achieves lower expected-rating MAE and RMSE, while the teacher has lower median-based MAE and RPS.
DuoLM also outperforms the rating priors, Content kNN, and raw Qwen3-0.6B across all reported average metrics.

\paragraph{Next-item recommendation.}
DuoLM achieves higher NDCG@10 than HLLM on Books (0.5586 vs.\ 0.5508) and Video Games (0.5298 vs.\ 0.5025), although HLLM achieves higher HR@1 on Books (Table~\ref{exp:recommendation-quality}).
Compared with the expert teacher model, DuoLM achieves higher NDCG@10 on Books (0.5586 vs.\ 0.5440) and comparable performance on Video Games (0.5298 vs.\ 0.5313).
Together with the rating prediction results, these findings show that DuoLM maintains quality comparable to that of the full-text teacher across the evaluated tasks and datasets.
\begin{table}[t]
\centering
\begingroup
\small
\definecolor{tableTwoGroup}{HTML}{F3F3F3}
\setlength{\tabcolsep}{3pt}
\renewcommand{\arraystretch}{1.0}

\caption{Next-item recommendation on Amazon Books and Amazon Video Games.}
\label{exp:recommendation-quality}

\vspace{-1em}
\begin{tabular*}{\columnwidth}{@{\extracolsep{\fill}}lrrr@{}}
    \toprule
    \textbf{Method}
    & \textbf{HR@1}
    & \textbf{HR@10}
    & \textbf{NDCG@10} \\
    \midrule
    \rowcolor{tableTwoGroup}[0pt][0pt]
    \multicolumn{4}{@{}l}{\textbf{Amazon Books}} \\
    \addlinespace[1pt]
    Popularity
    & 0.1770 & 0.4382 & 0.2944 \\
    HLLM
    & \textbf{0.4022} & 0.7168 & 0.5508 \\
    \addlinespace[1pt]
    FIE (m=1)
    & 0.3870 & \underline{0.7418} & \underline{0.5538} \\
    FIR (m=1)
    & \underline{0.3910} & \textbf{0.7464} & \textbf{0.5586} \\
    \addlinespace[1pt]
    Teacher model
    & 0.3824 & 0.7270 & 0.5440 \\
    \midrule
    \rowcolor{tableTwoGroup}[0pt][0pt]
    \multicolumn{4}{@{}l}{\textbf{Amazon Video Games}} \\
    \addlinespace[1pt]
    Popularity
    & 0.1460 & 0.5000 & 0.3012 \\
    History similarity
    & 0.0737 & 0.3057 & 0.1722 \\
    ContentSASRec
    & 0.1853 & 0.5970 & 0.3714 \\
    HLLM
    & 0.3057 & 0.7150 & 0.5025 \\
    \addlinespace[1pt]
    FIR (m=1)
    & \textbf{0.3240} & \underline{0.7650} & \underline{0.5298} \\
    \addlinespace[1pt]
    Teacher model
    & \underline{0.3237} & \textbf{0.7697} & \textbf{0.5313} \\
    \bottomrule
\end{tabular*}
\endgroup
\end{table}

\paragraph{RQ2: Does more user history improve personalized scoring quality?}
As $K$ increases from 1 to 10, DuoLM's NLL decreases from 0.9848 to 0.8806, its median-based MAE from 0.5705 to 0.5056, and its RPS from 0.1044 to 0.0911.
These results show that DuoLM benefits from additional history even with a single vector per item.
Complete results and supplementary within-user ranking analysis appear in Appendix~\ref{app:within-user}.

\subsection{End-to-End Serving Efficiency}
\label{sec:serving-efficiency}

\paragraph{RQ3: How fast can DuoLM serve personalized recommendations?}
We compare DuoLM with the teacher served on vLLM (with automatic prefix caching) and SGLang, using the same physical H100 80GB GPU and request trace with 100 candidates per request.

As concurrency increases from $C=1$ to $C=64$, DuoLM's throughput rises from approximately 8,870 to 26,400 candidate scores per second (Table~\ref{exp:recommendation-throughput}).
Across these settings, DuoLM achieves $22.7$--$60.1\times$ the throughput of the teacher on vLLM and $37.6$--$109.9\times$ that on SGLang.

These throughput gains accompany lower full-request p99 latency at every tested concurrency level.
At $C=64$, for example, DuoLM's p99 latency is 536.90\,ms, compared with 17,093.82\,ms for the teacher on vLLM.
Together, these results show that DuoLM achieves substantially higher throughput and lower latency under the evaluated workloads.

With the model fixed, MIS improves throughput by
4.81--29.3$\times$, with a further
1.06--3.54$\times$ from CUDA Graph
(Section~\ref{sec:runtime-attribution}).

\subsection{Representation and Training Analysis}
\label{sec:representation-training}
\label{sec:analysis}

This section analyze how item readout, feedback placement, distillation, and self-alignment affect quality. Readout and loss ablations here share a separately trained default model.

\paragraph{RQ4: How should content be read out from the Encoder for effective use by the Ranker?}
We compare direct EOS readout with two dedicated learnable readout mechanisms on Books.
All three readouts produce one vector per item and use the same scale-calibrated interface and joint training objective.
\textbf{LastToken} uses the Encoder's final-layer hidden state at EOS.
\textbf{MEM1} appends one learnable memory position after EOS and uses its final-layer hidden state as the item vector, following PISCO~\citep{louis-etal-2025-pisco} and CLaRa~\citep{he2025clarabridgingretrievalgeneration}.
\textbf{Perceiver1} aggregates the Encoder's output states with one learnable query, following Perceiver~\citep{jaegle2021perceiver} and Flamingo~\citep{alayrac2022flamingo}.

\begin{table}[t]
\centering
\begingroup
\fontsize{10pt}{12pt}\selectfont
\setlength{\tabcolsep}{3pt}
\renewcommand{\arraystretch}{1.12}
\caption{Single-vector readouts on Books.
Quality is evaluated on 5,000 users, covers 7,424 training updates.}
\label{exp:readout}
\begin{tabular*}{\columnwidth}{@{\extracolsep{\fill}}lrrr@{}}
\toprule
Metric & LastToken & MEM1 & Perceiver1 \\
\midrule
HR@1 $\uparrow$
& \textbf{0.4024} & \underline{0.3900} & 0.3588 \\
HR@10 $\uparrow$
& \textbf{0.7452} & \underline{0.7374} & 0.7320 \\
MRR@10 $\uparrow$
& \textbf{0.5049} & \underline{0.4982} & 0.4714 \\
NDCG@10 $\uparrow$
& \textbf{0.5621} & \underline{0.5555} & 0.5335 \\
\midrule
Added parameters
& 0 & 1,024 & 12,599,296 \\
Update time (h)
& 7.39 & 7.54 & 8.96 \\
\bottomrule
\end{tabular*}
\endgroup
\end{table}

LastToken achieves the highest scores across all four ranking metrics (Table~\ref{exp:readout}).
It also introduces no additional readout parameters and records the shortest total update time: 7.39 hours for 7,424 training updates, compared with 7.54 for MEM1 and 8.96 for Perceiver1.
We therefore adopt LastToken as the default readout.

\paragraph{RQ5: Should user feedback enter the Encoder or remain with the Ranker?}
FIE achieves slightly lower rating NLL (0.9236 vs.\ 0.9249), while FIR achieves higher Books recommendation NDCG@10 (0.5586 vs.\ 0.5538), as shown in Table~\ref{exp:feedback-placement}.
Given their comparable quality, we adopt FIR to reuse item representations independently of user feedback.
Config detailed in Appendix~\ref{app:protocol}.

\begin{table}[t]
\centering
\begingroup
\fontsize{10pt}{12pt}\selectfont
\setlength{\tabcolsep}{3pt}
\renewcommand{\arraystretch}{1.16}
\caption{
Comparison of FIE and FIR using separately trained
task recipes.
}
\label{exp:feedback-placement}
\begin{tabular*}{\columnwidth}{@{\extracolsep{\fill}}llrr@{}}
\toprule
Design & Reuse unit
& \shortstack{Rating\\NLL $\downarrow$}
& \shortstack{Books\\NDCG@10 $\uparrow$} \\
\midrule
FIE & Item--rating pair
& \textbf{0.9236} & \underline{0.5538} \\
FIR & Item
& \underline{0.9249} & \textbf{0.5586} \\
\bottomrule
\end{tabular*}
\endgroup
\end{table}

\begin{table}[t]
\centering
\begingroup
\fontsize{10pt}{12pt}\selectfont
\setlength{\tabcolsep}{3pt}
\renewcommand{\arraystretch}{1.16}
\caption{
Distillation study on Books ($m=1$).
Default CE+KD uses final KD weight $\beta=0.5$ and
temperature $\tau=1$.
Each variant changes only the indicated setting.
metrics are reported on all 5,000 users.
}
\label{tab:distillation}
\begin{tabular*}{\columnwidth}{@{\extracolsep{\fill}}lrrr@{}}
\toprule
Training setting & HR@1 & HR@10 & NDCG@10 \\
\midrule
CE only
& 0.3748 & \textbf{0.7474} & 0.5502 \\
CE+KD (default)
& \textbf{0.4024} & \underline{0.7452} & \textbf{0.5621} \\
KD weight $\beta=0.9$
& 0.3774 & 0.7284 & 0.5428 \\
Temperature $\tau=2$
& \underline{0.3904} & 0.7424 & \underline{0.5568} \\
\bottomrule
\end{tabular*}
\endgroup
\end{table}

\begin{table}[t]
\centering
\begingroup
\small
\setlength{\tabcolsep}{3pt}
\renewcommand{\arraystretch}{1.08}
\caption{End-to-end Books recommendation serving on one H100 with 100 candidates per request. $C$ denotes concurrency.}
\label{exp:recommendation-throughput}
\begin{tabular*}{\columnwidth}{@{\extracolsep{\fill}}crrr@{}}
\toprule
\multirow{2}{*}{$C$}
& \multicolumn{2}{c}{Teacher model}
& \multirow{2}{*}{DuoLM} \\
\cmidrule(lr){2-3}
& vLLM & SGLang & \\
\midrule
\multicolumn{4}{@{}l}{\textit{Throughput (candidate scores/s)} $\uparrow$} \\
\addlinespace[2pt]
1  & \underline{390.98} & 235.87 & \textbf{8\,870.36} \\
8  & \underline{438.98} & 241.55 & \textbf{20\,740.56} \\
32 & \underline{435.67} & 241.31 & \textbf{25\,506.11} \\
64 & \underline{439.08} & 240.29 & \textbf{26\,403.98} \\
\midrule
\multicolumn{4}{@{}l}{\textit{Full-request p99 latency (ms)} $\downarrow$} \\
\addlinespace[2pt]
1  & \underline{354.87}     & 706.46     & \textbf{12.35} \\
8  & \underline{2\,099.20}  & 4\,138.95  & \textbf{43.89} \\
32 & \underline{8\,155.82}  & 14\,701.82 & \textbf{389.27} \\
64 & \underline{17\,093.82} & 28\,189.04 & \textbf{536.90} \\
\bottomrule
\end{tabular*}
\endgroup
\end{table}

\Needspace{4\baselineskip}
\paragraph{RQ6: Does distillation from the full-text teacher help beyond recommendation labels alone?}
We compare label-only training (CE-only) with the default CE+KD objective.
On Books, adding distillation improves HR@1 from 0.3748 to 0.4024 and NDCG@10 from 0.5502 to 0.5621, while HR@10 decreases slightly from 0.7474 to 0.7452 (Table~\ref{tab:distillation}).
These results indicate that teacher supervision complements recommendation labels, supporting CE+KD as default training objective.

\begin{table}[t]
\centering
\begingroup
\fontsize{10pt}{12pt}\selectfont
\setlength{\tabcolsep}{3pt}
\renewcommand{\arraystretch}{1.16}
\caption{
Self-alignment on Books ($m=1$).
Both rows use default KD.
Alignment jointly adds output KL and hidden-state cosine losses.
}
\label{tab:self-alignment}
\begin{tabular*}{\columnwidth}{@{\extracolsep{\fill}}lrrr@{}}
\toprule
Self-alignment & HR@1 & HR@10 & NDCG@10 \\
\midrule
Without (default)
& \textbf{0.4024} & \textbf{0.7452} & \textbf{0.5621} \\
With
& 0.3900 & 0.7414 & 0.5554 \\
\bottomrule
\end{tabular*}
\endgroup
\end{table}

\paragraph{RQ7: Does a higher distillation weight or temperature improve recommendation quality?}
We separately increase the final distillation weight and temperature from their default values.
On Books, raising $\beta$ from 0.5 to 0.9 reduces NDCG@10 from 0.5621 to 0.5428, while raising $\tau$ from 1 to 2 yields 0.5568 (Table~\ref{tab:distillation}).
Neither change improves upon the default, so we retain default.

\begingroup
\widowpenalty=10000
\paragraph{RQ8: Is additional full-text self-alignment worthwhile?}
We add self-alignment to the default CE+KD objective, encouraging consistency between the same Ranker's outputs and hidden representations under full-text and compressed inputs.
On Books, NDCG@10 decreases from 0.5621 to 0.5554 (Table~\ref{tab:self-alignment}), while cumulative training update time increases from 7.39 to 16.95 hours, or $2.29\times$ (Appendix~\ref{app:training}).
We therefore omit self-alignment from the default configuration, as it increases cost without improving recommendation quality.
\par\endgroup

\subsection{Runtime Attribution and Workload Analysis}
\label{sec:runtime-attribution}
\label{sec:runtime-analysis}

\paragraph{RQ9: Which runtime designs contribute to the speedup, and how do the gains vary with workload?}

\begin{figure}[t]
  \centering
  \includegraphics[width=0.90\linewidth]{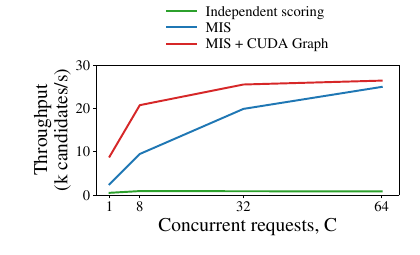}
  \caption{Execution ablation with 100 candidates per request.}
  \label{fig:execution-ablation}
\end{figure}

\paragraph{MIS and CUDA Graph.}
With the recommendation model and request trace fixed,
MIS improves throughput over independent scoring by
$4.81\times$ at $C=1$ and $29.3\times$ at $C=64$
(Figure~\ref{fig:execution-ablation} and Table~\ref{exp:runtime}).
CUDA Graph provides a further $3.54\times$ speedup at
$C=1$, compared with $1.06\times$ at $C=64$.
MIS therefore yields larger relative gains at higher
concurrency, while graph replay contributes more at
lower concurrency.

\begin{figure}[t]
  \centering
  \includegraphics[width=\linewidth]{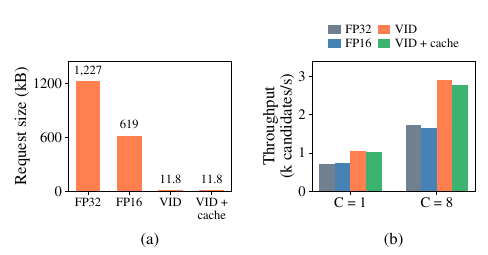}
  \caption{VID input-path diagnostics: request size and throughput.}
  \label{fig:vid-input-path}
\end{figure}

\paragraph{Virtual ID inputs.}
In the FIR rating diagnostic ($m=8$), VID reduces
request size by 99.0\% and increases throughput by
66.5\% at $C=8$, relative to transferring FP32 vectors
(Figure~\ref{fig:vid-input-path}).
FP16 transfer reduces request size but does not
consistently improve throughput, indicating that
fewer transferred bytes alone do not guarantee
faster serving.

\paragraph{CPU input preparation.}
At an arrival rate of 100 requests/s, increasing
input-compilation processes from one to two brings
the single-vector rating service close to the offered
request rate, showing that CPU input preparation can
also limit end-to-end throughput.
Detailed results appear in Appendix~\ref{app:serving}.

\section{Related Work}
\label{sec:related-work}

\paragraph{Language models for personalized recommendation.}
In traditional recommendation pipelines, two-tower models encode users and items separately for efficient retrieval~\citep{Yi_2019}.
For models that rely on ID embeddings, sparse or missing interaction data limit representation learning for long-tail and cold-start items~\citep{li2023textneedlearninglanguage}.
Language-based approaches use pretrained models to encode item content and model user preferences~\citep{zhang2025collm,liao2024llara}, including hierarchical item and user modeling in HLLM~\citep{chen2024hllmenhancingsequentialrecommendations}. Whereas HLLM scores candidates by user--item representation
similarity, DuoLM lets each candidate attend directly to
individual history items and their associated feedback.

\paragraph{Personalization and compact representations.}
Prior work personalizes language models through history representations~\citep{ning2024userllmefficientllmcontextualization,liu-etal-2025-llms}, and decoding adjustments~\citep{song2026cardclusterleveladaptationrewardguided}.
Continuous context compression replaces long text with compact representations consumed by downstream LLM~\citep{NEURIPS2024_c5cf13bf,louis-etal-2025-pisco,he2025clarabridgingretrievalgeneration,mu2023gist}.

\paragraph{Efficient language model inference.}
KV-cache management and shared-prefix execution reduce redundant storage and computation~\citep{kwon2023efficient,NEURIPS2024_724be447,ye-etal-2024-chunkattention}.
Optimized attention kernels reduce memory traffic~\citep{NEURIPS2022_67d57c32}, and runtimes support CUDA Graph capture and replay~\citep{nvidiaCudaGraphs}.

\section{Conclusion}
\label{sec:conclusion}

We presented DuoLM, a framework that jointly trains two LLMs
for reusable item representations and personalized recommendation,
paired with an efficient serving runtime.
DuoLM achieves recommendation quality comparable to the teacher.
For Books recommendation on a single H100, it delivers up to
$60.1\times$ the throughput of the teacher served by vLLM.

\section*{Limitations}

Our evaluation focuses on Amazon Books and Video Games with 0.6B language-model backbones. Results on these datasets may not generalize to other domains, larger models, or recommendation settings with rapidly changing user preferences. Recommendation quality is evaluated on fixed candidate sets containing randomly sampled unobserved items, which may be easier to distinguish than candidates produced by a strong retrieval system.

Compressing each item into a single vector may discard details relevant to particular users. Offline representations also require regeneration when item content or encoder parameters change. Our throughput measurements cover online serving and exclude training and offline encoding costs. Finally, the serving results are measured on an H100 under controlled request loads. Performance under different hardware, traffic patterns, and latency constraints requires further evaluation.

\ifreferencesready
  \bibliography{references}

\begin{thebibliography}{41}
\providecommand{\natexlab}[1]{#1}

\bibitem[{Alayrac et~al.(2022)Alayrac, Donahue, Luc, Miech, Barr, Hasson, Lenc,
  Mensch, Millican, Reynolds, Ring, Rutherford, Cabi, Han, Gong, Samangooei,
  Monteiro, Menick, Borgeaud, Brock, Nematzadeh, Sharifzadeh, Binkowski,
  Barreira, Vinyals, Zisserman, and Simonyan}]{alayrac2022flamingo}
Jean-Baptiste Alayrac, Jeff Donahue, Pauline Luc, Antoine Miech, Iain Barr,
  Yana Hasson, Karel Lenc, Arthur Mensch, Katie Millican, Malcolm Reynolds,
  Roman Ring, Eliza Rutherford, Serkan Cabi, Tengda Han, Zhitao Gong, Sina
  Samangooei, Marianne Monteiro, Jacob Menick, Sebastian Borgeaud, and 8
  others. 2022.
\newblock \href {https://arxiv.org/abs/2204.14198} {{Flamingo: a Visual
  Language Model for Few-Shot Learning}}.
\newblock \emph{Preprint}, arXiv:2204.14198.

\bibitem[{Bao et~al.(2023)Bao, Zhang, Zhang, Wang, Feng, and
  He}]{bao2023tallrec}
Keqin Bao, Jizhi Zhang, Yang Zhang, Wenjie Wang, Fuli Feng, and Xiangnan He.
  2023.
\newblock \href {https://doi.org/10.1145/3604915.3608857} {{TALLRec}: An
  effective and efficient tuning framework to align large language model with
  recommendation}.
\newblock In \emph{Proceedings of the 17th ACM Conference on Recommender
  Systems}, RecSys '23, pages 1007--1014. Association for Computing Machinery.

\bibitem[{Chen et~al.(2024)Chen, Chi, Peng, and
  Yuan}]{chen2024hllmenhancingsequentialrecommendations}
Junyi Chen, Lu~Chi, Bingyue Peng, and Zehuan Yuan. 2024.
\newblock \href {https://arxiv.org/abs/2409.12740} {Hllm: Enhancing sequential
  recommendations via hierarchical large language models for item and user
  modeling}.
\newblock \emph{Preprint}, arXiv:2409.12740.

\bibitem[{Chen et~al.(2016)Chen, Xu, Zhang, and Guestrin}]{chen2016training}
Tianqi Chen, Bing Xu, Chiyuan Zhang, and Carlos Guestrin. 2016.
\newblock \href {https://arxiv.org/abs/1604.06174} {Training deep nets with
  sublinear memory cost}.
\newblock \emph{Preprint}, arXiv:1604.06174.

\bibitem[{Cheng et~al.(2024)Cheng, Wang, Zhang, Ge, Chen, Wei, Zhang, and
  Zhao}]{NEURIPS2024_c5cf13bf}
Xin Cheng, Xun Wang, Xingxing Zhang, Tao Ge, Si-Qing Chen, Furu Wei, Huishuai
  Zhang, and Dongyan Zhao. 2024.
\newblock \href {https://doi.org/10.52202/079017-3476} {xrag: Extreme context
  compression for retrieval-augmented generation with one token}.
\newblock In \emph{Advances in Neural Information Processing Systems},
  volume~37, pages 109487--109516. Curran Associates, Inc.

\bibitem[{Chevalier et~al.(2023)Chevalier, Wettig, Ajith, and
  Chen}]{chevalier2023autocompressors}
Alexis Chevalier, Alexander Wettig, Anirudh Ajith, and Danqi Chen. 2023.
\newblock \href {https://doi.org/10.18653/v1/2023.emnlp-main.232} {Adapting
  language models to compress contexts}.
\newblock In \emph{Proceedings of the 2023 Conference on Empirical Methods in
  Natural Language Processing}, pages 3829--3846, Singapore. Association for
  Computational Linguistics.

\bibitem[{Dao et~al.(2022)Dao, Fu, Ermon, Rudra, and
  R\'{e}}]{NEURIPS2022_67d57c32}
Tri Dao, Dan Fu, Stefano Ermon, Atri Rudra, and Christopher R\'{e}. 2022.
\newblock \href {https://doi.org/10.52202/068431-1189} {Flashattention: Fast
  and memory-efficient exact attention with io-awareness}.
\newblock In \emph{Advances in Neural Information Processing Systems},
  volume~35, pages 16344--16359. Curran Associates, Inc.

\bibitem[{Efron(1979)}]{efron1979bootstrap}
Bradley Efron. 1979.
\newblock \href {https://doi.org/10.1214/aos/1176344552} {Bootstrap methods:
  Another look at the jackknife}.
\newblock \emph{The Annals of Statistics}, 7(1):1--26.

\bibitem[{Ge et~al.(2024)Ge, Hu, Wang, Wang, Chen, and
  Wei}]{ge2024incontextautoencodercontextcompression}
Tao Ge, Jing Hu, Lei Wang, Xun Wang, Si-Qing Chen, and Furu Wei. 2024.
\newblock \href {https://arxiv.org/abs/2307.06945} {In-context autoencoder for
  context compression in a large language model}.
\newblock \emph{Preprint}, arXiv:2307.06945.

\bibitem[{Geng et~al.(2022)Geng, Liu, Fu, Ge, and Zhang}]{geng2022p5}
Shijie Geng, Shuchang Liu, Zuohui Fu, Yingqiang Ge, and Yongfeng Zhang. 2022.
\newblock \href {https://doi.org/10.1145/3523227.3546767} {Recommendation as
  language processing ({RLP}): A unified pretrain, personalized prompt \&
  predict paradigm ({P5})}.
\newblock In \emph{Proceedings of the 16th ACM Conference on Recommender
  Systems}, pages 299--315. Association for Computing Machinery.

\bibitem[{Gneiting and Raftery(2007)}]{gneiting2007strictly}
Tilmann Gneiting and Adrian~E. Raftery. 2007.
\newblock \href {https://doi.org/10.1198/016214506000001437} {Strictly proper
  scoring rules, prediction, and estimation}.
\newblock \emph{Journal of the American Statistical Association},
  102(477):359--378.

\bibitem[{He et~al.(2025)He, Bai, Williamson, Pan, Jaitly, and
  Zhang}]{he2025clarabridgingretrievalgeneration}
Jie He, Richard~He Bai, Sinead Williamson, Jeff~Z. Pan, Navdeep Jaitly, and
  Yizhe Zhang. 2025.
\newblock \href {https://arxiv.org/abs/2511.18659} {Clara: Bridging retrieval
  and generation with continuous latent reasoning}.
\newblock \emph{Preprint}, arXiv:2511.18659.

\bibitem[{Hinton et~al.(2015)Hinton, Vinyals, and
  Dean}]{hinton2015distillingknowledgeneuralnetwork}
Geoffrey Hinton, Oriol Vinyals, and Jeff Dean. 2015.
\newblock \href {https://arxiv.org/abs/1503.02531} {Distilling the knowledge in
  a neural network}.
\newblock \emph{Preprint}, arXiv:1503.02531.

\bibitem[{Hou et~al.(2024)Hou, Li, He, Yan, Chen, and
  McAuley}]{hou2024bridging}
Yupeng Hou, Jiacheng Li, Zhankui He, An~Yan, Xiusi Chen, and Julian McAuley.
  2024.
\newblock \href {https://arxiv.org/abs/2403.03952} {Bridging language and items
  for retrieval and recommendation}.
\newblock \emph{arXiv preprint arXiv:2403.03952}.

\bibitem[{Hou et~al.(2022)Hou, Mu, Zhao, Li, Ding, and
  Wen}]{hou2022universalsequencerepresentationlearning}
Yupeng Hou, Shanlei Mu, Wayne~Xin Zhao, Yaliang Li, Bolin Ding, and Ji-Rong
  Wen. 2022.
\newblock \href {https://arxiv.org/abs/2206.05941} {Towards universal sequence
  representation learning for recommender systems}.
\newblock \emph{Preprint}, arXiv:2206.05941.

\bibitem[{Jaegle et~al.(2021)Jaegle, Gimeno, Brock, Zisserman, Vinyals, and
  Carreira}]{jaegle2021perceiver}
Andrew Jaegle, Felix Gimeno, Andrew Brock, Andrew Zisserman, Oriol Vinyals, and
  Joao Carreira. 2021.
\newblock \href {https://arxiv.org/abs/2103.03206} {{Perceiver: General
  Perception with Iterative Attention}}.
\newblock \emph{Preprint}, arXiv:2103.03206.

\bibitem[{J{\"a}rvelin and Kek{\"a}l{\"a}inen(2002)}]{jarvelin2002cumulated}
Kalervo J{\"a}rvelin and Jaana Kek{\"a}l{\"a}inen. 2002.
\newblock \href {https://doi.org/10.1145/582415.582418} {Cumulated gain-based
  evaluation of {IR} techniques}.
\newblock \emph{ACM Transactions on Information Systems}, 20(4):422--446.

\bibitem[{Juravsky et~al.(2024)Juravsky, Brown, Ehrlich, Fu, Ré, and
  Mirhoseini}]{juravsky2024hydragen}
Jordan Juravsky, Bradley Brown, Ryan Ehrlich, Daniel~Y. Fu, Christopher Ré,
  and Azalia Mirhoseini. 2024.
\newblock \href {https://arxiv.org/abs/2402.05099} {Hydragen: High-throughput
  llm inference with shared prefixes}.
\newblock \emph{Preprint}, arXiv:2402.05099.

\bibitem[{Kang and McAuley(2018)}]{kang2018sasrec}
Wang-Cheng Kang and Julian McAuley. 2018.
\newblock \href {https://arxiv.org/abs/1808.09781} {{Self-Attentive Sequential
  Recommendation}}.
\newblock \emph{Preprint}, arXiv:1808.09781.

\bibitem[{Kim et~al.(2024)Kim, Kang, Choi, Kim, Yang, and
  Park}]{kim2024allmrec}
Sein Kim, Hongseok Kang, Seungyoon Choi, Donghyun Kim, Minchul Yang, and
  Chanyoung Park. 2024.
\newblock \href {https://doi.org/10.1145/3637528.3671931} {Large language
  models meet collaborative filtering: An efficient all-round {LLM}-based
  recommender system}.
\newblock In \emph{Proceedings of the 30th ACM SIGKDD Conference on Knowledge
  Discovery and Data Mining}, KDD '24, pages 1395--1406. Association for
  Computing Machinery.

\bibitem[{Kwon et~al.(2023)Kwon, Li, Zhuang, Sheng, Zheng, Yu, Gonzalez, Zhang,
  and Stoica}]{kwon2023efficient}
Woosuk Kwon, Zhuohan Li, Siyuan Zhuang, Ying Sheng, Lianmin Zheng, Cody~Hao Yu,
  Joseph~E. Gonzalez, Hao Zhang, and Ion Stoica. 2023.
\newblock Efficient memory management for large language model serving with
  pagedattention.
\newblock In \emph{Proceedings of the ACM SIGOPS 29th Symposium on Operating
  Systems Principles}.

\bibitem[{Li et~al.(2023)Li, Wang, Li, Fu, Shen, Shang, and
  McAuley}]{li2023textneedlearninglanguage}
Jiacheng Li, Ming Wang, Jin Li, Jinmiao Fu, Xin Shen, Jingbo Shang, and Julian
  McAuley. 2023.
\newblock \href {https://arxiv.org/abs/2305.13731} {Text is all you need:
  Learning language representations for sequential recommendation}.
\newblock \emph{Preprint}, arXiv:2305.13731.

\bibitem[{Liao et~al.(2024)Liao, Li, Yang, Wu, Yuan, Wang, and
  He}]{liao2024llara}
Jiayi Liao, Sihang Li, Zhengyi Yang, Jiancan Wu, Yancheng Yuan, Xiang Wang, and
  Xiangnan He. 2024.
\newblock \href {https://doi.org/10.1145/3626772.3657690} {{LLaRA}: Large
  language-recommendation assistant}.
\newblock In \emph{Proceedings of the 47th International ACM SIGIR Conference
  on Research and Development in Information Retrieval}, pages 1785--1795.
  Association for Computing Machinery.

\bibitem[{Liu et~al.(2025)Liu, Zhu, Wang, Wei, Min, Lu, Wang, Yin, and
  Dou}]{liu-etal-2025-llms}
Jiongnan Liu, Yutao Zhu, Shuting Wang, Xiaochi Wei, Erxue Min, Yu~Lu,
  Shuaiqiang Wang, Dawei Yin, and Zhicheng Dou. 2025.
\newblock \href {https://doi.org/10.18653/v1/2025.acl-long.461} {{LLM}s +
  persona-plug = personalized {LLM}s}.
\newblock In \emph{Proceedings of the 63rd Annual Meeting of the Association
  for Computational Linguistics (Volume 1: Long Papers)}, pages 9373--9385,
  Vienna, Austria. Association for Computational Linguistics.

\bibitem[{Louis et~al.(2025)Louis, D{\'e}jean, and
  Clinchant}]{louis-etal-2025-pisco}
Maxime Louis, Herv{\'e} D{\'e}jean, and St{\'e}phane Clinchant. 2025.
\newblock \href {https://doi.org/10.18653/v1/2025.findings-acl.800} {{PISCO}:
  Pretty simple compression for retrieval-augmented generation}.
\newblock In \emph{Findings of the Association for Computational Linguistics:
  ACL 2025}, pages 15506--15521, Vienna, Austria. Association for Computational
  Linguistics.

\bibitem[{McSherry and Najork(2008)}]{mcsherry2008tied}
Frank McSherry and Marc Najork. 2008.
\newblock \href {https://doi.org/10.1007/978-3-540-78646-7_38} {Computing
  information retrieval performance measures efficiently in the presence of
  tied scores}.
\newblock In \emph{Advances in Information Retrieval}, volume 4956 of
  \emph{Lecture Notes in Computer Science}, pages 414--421. Springer.

\bibitem[{Mu et~al.(2023)Mu, Li, and Goodman}]{mu2023gist}
Jesse Mu, Xiang Li, and Noah Goodman. 2023.
\newblock \href {https://doi.org/10.52202/075280-0848} {Learning to compress
  prompts with gist tokens}.
\newblock In \emph{Advances in Neural Information Processing Systems},
  volume~36, pages 19327--19352.

\bibitem[{Ning et~al.(2024)Ning, Liu, Wu, Wu, Berlowitz, Prakash, Green,
  O'Banion, and Xie}]{ning2024userllmefficientllmcontextualization}
Lin Ning, Luyang Liu, Jiaxing Wu, Neo Wu, Devora Berlowitz, Sushant Prakash,
  Bradley Green, Shawn O'Banion, and Jun Xie. 2024.
\newblock \href {https://arxiv.org/abs/2402.13598} {User-llm: Efficient llm
  contextualization with user embeddings}.
\newblock \emph{Preprint}, arXiv:2402.13598.

\bibitem[{{NVIDIA Corporation}(2026)}]{nvidiaCudaGraphs}
{NVIDIA Corporation}. 2026.
\newblock \href
  {https://docs.nvidia.com/cuda/cuda-programming-guide/04-special-topics/cuda-graphs.html}
  {\emph{{CUDA Programming Guide}: {CUDA Graphs}}}.
\newblock NVIDIA Corporation.
\newblock Section 4.2. Last updated September 10, 2026. Accessed October 9,
  2026.

\bibitem[{{Qwen Team}(2025)}]{qwen2025qwen3}
{Qwen Team}. 2025.
\newblock \href {https://arxiv.org/abs/2505.09388} {{Qwen3} technical report}.
\newblock \emph{Preprint}, arXiv:2505.09388.

\bibitem[{Sanh et~al.(2019)Sanh, Debut, Chaumond, and
  Wolf}]{sanh2019distilbert}
Victor Sanh, Lysandre Debut, Julien Chaumond, and Thomas Wolf. 2019.
\newblock \href {https://arxiv.org/abs/1910.01108} {{DistilBERT}, a distilled
  version of {BERT}: Smaller, faster, cheaper and lighter}.
\newblock \emph{Preprint}, arXiv:1910.01108.

\bibitem[{Song et~al.(2026)Song, Wu, Zhang, Shen, Yuan, Lu, Wang, Wang, Dutt,
  and Rahmani}]{song2026cardclusterleveladaptationrewardguided}
Yutong Song, Jiang Wu, Weijia Zhang, Chengze Shen, Shaofan Yuan, Weitao Lu,
  Jian Wang, Yu~Wang, Nikil Dutt, and Amir~M. Rahmani. 2026.
\newblock \href {https://arxiv.org/abs/2601.06352} {Card: Cluster-level
  adaptation with reward-guided decoding for personalized text generation}.
\newblock \emph{Preprint}, arXiv:2601.06352.

\bibitem[{Vaswani et~al.(2017)Vaswani, Shazeer, Parmar, Uszkoreit, Jones,
  Gomez, Kaiser, and Polosukhin}]{NIPS2017_3f5ee243}
Ashish Vaswani, Noam Shazeer, Niki Parmar, Jakob Uszkoreit, Llion Jones,
  Aidan~N. Gomez, {\L}ukasz Kaiser, and Illia Polosukhin. 2017.
\newblock \href
  {https://proceedings.neurips.cc/paper_files/paper/2017/file/3f5ee243547dee91fbd053c1c4a845aa-Paper.pdf}
  {Attention is all you need}.
\newblock In \emph{Advances in Neural Information Processing Systems},
  volume~30, pages 5998--6008.

\bibitem[{Ye et~al.(2024)Ye, Tao, Huang, and Li}]{ye-etal-2024-chunkattention}
Lu~Ye, Ze~Tao, Yong Huang, and Yang Li. 2024.
\newblock \href {https://doi.org/10.18653/v1/2024.acl-long.623}
  {{ChunkAttention}: Efficient self-attention with prefix-aware {KV} cache and
  two-phase partition}.
\newblock In \emph{Proceedings of the 62nd Annual Meeting of the Association
  for Computational Linguistics (Volume 1: Long Papers)}, pages 11608--11620,
  Bangkok, Thailand. Association for Computational Linguistics.

\bibitem[{Ye et~al.(2025)Ye, Chen, Lai, Lin, Zhang, Wang, Chen, Kasikci,
  Grover, Krishnamurthy, and Ceze}]{MLSYS2025_dbf02b21}
Zihao Ye, Lequn Chen, Ruihang Lai, Wuwei Lin, Yineng Zhang, Stephanie Wang,
  Tianqi Chen, Baris Kasikci, Vinod Grover, Arvind Krishnamurthy, and Luis
  Ceze. 2025.
\newblock \href
  {https://proceedings.mlsys.org/paper_files/paper/2025/file/dbf02b21d77409a2db30e56866a8ab3a-Paper-Conference.pdf}
  {Flashinfer: Efficient and customizable attention engine for llm inference
  serving}.
\newblock In \emph{Proceedings of Machine Learning and Systems}, volume~7.
  MLSys.

\bibitem[{Yi et~al.(2019)Yi, Yang, Hong, Cheng, Heldt, Kumthekar, Zhao, Wei,
  and Chi}]{Yi_2019}
Xinyang Yi, Ji~Yang, Lichan Hong, Derek~Zhiyuan Cheng, Lukasz Heldt, Aditee
  Kumthekar, Zhe Zhao, Li~Wei, and Ed~Chi. 2019.
\newblock \href {https://doi.org/10.1145/3298689.3346996}
  {Sampling-bias-corrected neural modeling for large corpus item
  recommendations}.
\newblock In \emph{Proceedings of the 13th ACM Conference on Recommender
  Systems}, RecSys '19, pages 269--277. ACM.

\bibitem[{Yue et~al.(2023)Yue, Rabhi, Moreira, Wang, and
  Oldridge}]{yue2023llamarec}
Zhenrui Yue, Sara Rabhi, Gabriel de Souza~Pereira Moreira, Dong Wang, and Even
  Oldridge. 2023.
\newblock \href {https://arxiv.org/abs/2311.02089} {{LlamaRec}: Two-stage
  recommendation using large language models for ranking}.
\newblock \emph{Preprint}, arXiv:2311.02089.
\newblock Presented at the Personalized Generative AI Workshop at CIKM 2023.

\bibitem[{Zhang and Sennrich(2019)}]{zhang2019root}
Biao Zhang and Rico Sennrich. 2019.
\newblock \href
  {https://papers.nips.cc/paper_files/paper/2019/hash/1e8a19426224ca89e83cef47f1e7f53b-Abstract.html}
  {Root mean square layer normalization}.
\newblock In \emph{Advances in Neural Information Processing Systems},
  volume~32.

\bibitem[{Zhang et~al.(2025{\natexlab{a}})Zhang, Feng, Zhang, Bao, Wang, and
  He}]{zhang2025collm}
Yang Zhang, Fuli Feng, Jizhi Zhang, Keqin Bao, Qifan Wang, and Xiangnan He.
  2025{\natexlab{a}}.
\newblock \href {https://doi.org/10.1109/TKDE.2025.3540912} {{CoLLM}:
  Integrating collaborative embeddings into large language models for
  recommendation}.
\newblock \emph{IEEE Transactions on Knowledge and Data Engineering},
  37(5):2329--2340.

\bibitem[{Zhang et~al.(2025{\natexlab{b}})Zhang, Li, Long, Zhang, Lin, Yang,
  Xie, Yang, Liu, Lin, Huang, and Zhou}]{zhang2025qwen3embedding}
Yanzhao Zhang, Mingxin Li, Dingkun Long, Xin Zhang, Huan Lin, Baosong Yang,
  Pengjun Xie, An~Yang, Dayiheng Liu, Junyang Lin, Fei Huang, and Jingren Zhou.
  2025{\natexlab{b}}.
\newblock \href {https://arxiv.org/abs/2506.05176} {{Qwen3 Embedding}:
  Advancing text embedding and reranking through foundation models}.
\newblock \emph{arXiv preprint arXiv:2506.05176}.

\bibitem[{Zheng et~al.(2024)Zheng, Yin, Xie, Sun, Huang, Yu, Cao, Kozyrakis,
  Stoica, Gonzalez, Barrett, and Sheng}]{NEURIPS2024_724be447}
Lianmin Zheng, Liangsheng Yin, Zhiqiang Xie, Chuyue Sun, Jeff Huang, Cody~Hao
  Yu, Shiyi Cao, Christos Kozyrakis, Ion Stoica, Joseph~E. Gonzalez, Clark
  Barrett, and Ying Sheng. 2024.
\newblock \href {https://doi.org/10.52202/079017-2000} {Sglang: Efficient
  execution of structured language model programs}.
\newblock In \emph{Advances in Neural Information Processing Systems},
  volume~37, pages 62557--62583. Curran Associates, Inc.

\end{thebibliography}
\else
  \section*{References}
\fi

\clearpage
\appendix
\section*{Appendix}
\makeatletter
\if@twocolumn\else\twocolumn\fi
\newcommand{\appFixedTable}[1]{%
  \par\noindent\begin{minipage}{\linewidth}%
  \def\@captype{table}\captionsetup{hypcap=false,skip=4pt}%
  #1\par\end{minipage}\par\vspace{2pt}}
\newcommand{\appFixedFigure}[1]{%
  \par\noindent\begin{minipage}{\linewidth}%
  \def\@captype{figure}\captionsetup{hypcap=false,skip=4pt}%
  #1\par\end{minipage}\par\vspace{2pt}}
\newcommand{\appWideTable}[1]{%
  \begin{table*}[!t]\captionsetup{skip=4pt}#1\end{table*}}
\newcommand{\appWideFigure}[1]{%
  \begin{figure*}[t]\captionsetup{skip=4pt}#1\end{figure*}}
\newcommand{\appColumnTable}[1]{%
  \begin{table}[!htbp]\captionsetup{skip=4pt}#1\end{table}}
\newcommand{\appColumnFigure}[1]{%
  \begin{figure}[!htbp]\captionsetup{skip=4pt}#1\end{figure}}
\makeatother

\raggedbottom
\setlength{\textfloatsep}{8pt plus 2pt minus 2pt}
\setlength{\floatsep}{8pt plus 2pt minus 2pt}
\setlength{\intextsep}{8pt plus 2pt minus 2pt}
\renewcommand{\thetable}{\thesection\arabic{table}}
\renewcommand{\theHtable}{\thesection.\arabic{table}}

\section{Experimental Protocol and Implementation}
\label{app:protocol}

This section provides the experimental details for rating prediction and next-item recommendation. It covers data construction and evaluation, model and baseline implementations, and serving measurements.

\subsection{Data Splits and Evaluation}
\label{app:data-eval}

We use the 5-core interactions and item metadata from Amazon Reviews 2023~\citep{hou2024bridging}, with splits determined by a fixed user hash. Item text includes titles, categories, features, and descriptions, and excludes target reviews, target ratings, and aggregate ratings.

\paragraph{Rating prediction.}
We use 10,000 training users and 1,000 development users. Each user's earliest ten valid items form a support pool, and targets occur after this pool. We evaluate histories of length $K\in\{1,3,5,10\}$. These histories are nested subsets of the support pool, selected in a rating-independent hash order and then presented chronologically. The development set contains 8,757 targets, yielding 35,028 predictions across the four values of $K$.

Rating metrics are first averaged within each user, then macro-averaged across users, and finally averaged with equal weight across $K$. MAE$_{\mathrm{med}}$ uses the first star rating at which the cumulative probability reaches or exceeds 0.5. MAE$_{\mathrm{exp}}$ and RMSE use the expected rating, $\hat y=\sum_{r=1}^{5}r p(r)$, where $p(r)$ is the predicted probability of rating $r$. RMSE takes the square root of the mean squared error within each user before averaging across users and then across $K$. For a true rating $y$, RPS~\citep{gneiting2007strictly} is defined as
\[
\mathrm{RPS}=\frac{1}{4}\sum_{r=1}^{4}
\left[F(r)-\mathbf{1}(y\leq r)\right]^2,
\]
where $F(r)$ is the cumulative probability of the predicted rating distribution.

\paragraph{Next-item recommendation.}
Books and Video Games each use 50,000 training users, yielding 237,558 and 233,868 training queries, respectively. Their item catalogs contain 495,063 and 25,611 items, respectively. Each history contains at most ten items, and each user contributes at most five training queries.

During training, each target is paired with seven negatives: four sampled uniformly and three retrieved using the history content. Evaluation uses one target and 99 fixed, randomly sampled items with which the user has not interacted. Users who cannot be scored remain in the macro-average denominator and contribute zero; there are seven such users in Books and six in Video Games.

\paragraph{Model selection and reporting.}
We select rating models by NLL averaged equally across $K$ on the full development set and report their results on the same set. Recommendation students are selected on a fixed development subset of 2,000 users. Books results use the full set of 5,000 users, which includes this subset; Video Games results use the remaining 3,000 users. The Books full-text teacher retains the checkpoint selected under the retrieved-candidate protocol and is then evaluated on the fixed 100-candidate protocol, comprising one target and 99 randomly sampled items with which the user has not interacted.

\subsection{Model and Baseline Implementations}
\label{app:models}

\paragraph{Model configuration.}
Encoder $A$ is initialized from Qwen3-Embedding-0.6B~\citep{zhang2025qwen3embedding}. The full-text teacher and ranker $B$ use Qwen3-0.6B-Base for recommendation and Qwen3-0.6B for rating prediction~\citep{qwen2025qwen3}. $B$ is initialized from the public base weights. Both $A$ and $B$ have 28 layers and a hidden dimension of 1,024.

Recommendation uses LastToken and a scalar scoring head; rating prediction uses Resampler/RMS and a five-star output head. For the RMS reference in rating prediction, we use the mean of the per-token RMS values. The main configuration uses FIR, $m=1$, and the short input-suffix template. Appendix~\ref{app:tail-template} compares the long- and short input-suffix templates.

Item content is capped at 512 tokens for recommendation and 256 tokens for rating prediction, measured with $B$'s tokenizer; $A$ additionally receives the template prefix and EOS. The rating FIE and FIR configurations in Table~\ref{exp:feedback-placement} use matched total update budgets and ranker input budgets, while the allocation of history vectors varies by recipe.

\paragraph{Statistical and content baselines.}
For rating prediction, Global prior uses the training rating distribution; Item prior uses item rating counts; and History prior uses rating counts from the current user's visible history. The latter two are smoothed toward the global distribution. Content kNN weights historical ratings by the similarity of frozen content vectors and follows the same smoothing approach. Smoothing and kNN parameters are selected by development NLL averaged equally across $K$ and then fixed.

For recommendation, Popularity counts the distinct users who interacted with each item in the union of training histories and targets. History similarity compares each candidate with the equally weighted mean of the user's history content vectors.

\paragraph{ContentSASRec/publicA.}
This content-based SASRec~\citep{kang2018sasrec} uses frozen Qwen3-Embedding-0.6B item vectors and trains a projection from 1,024 to 256 dimensions, rating and position embeddings, and a two-layer, four-head sequential Transformer. It scores candidates by the dot product of normalized user and candidate representations.

\paragraph{HLLM and HLLM-rating.}
For recommendation, HLLM~\citep{chen2024hllmenhancingsequentialrecommendations} uses two Qwen3 models of approximately 0.6B parameters to encode item content and user history, respectively. It ranks candidates by the cosine similarity between user and candidate representations, with the item sequence as the history input. Our adaptation for five-star prediction, HLLM-rating, adds star-rating embeddings to history item representations and feeds interaction features between the user and target representations into a classification network. The HLLM and HLLM-rating baselines are initialized from public weights and trained on their respective task data. HLLM-rating uses cross-entropy loss on true star ratings.

\paragraph{Full-text language models.}
Raw Qwen3-0.6B directly reads the user history, feedback, and target item text to predict a five-star rating. The full-text teacher is trained as described in Section~\ref{sec:learning}. It uses Qwen3's causal Transformer to jointly process the history, feedback, and original candidate text, then outputs candidate scores or probabilities over the five rating levels through an output head, without generating recommendation text.

\subsection{Serving Measurements}
\label{app:serving-protocol}

\paragraph{Timing scope.}
Throughput includes online input preparation, submission, transfer, queuing, model execution, and postprocessing. Offline encoding, startup, warmup, candidate retrieval, and wide-area network transfer are outside the timing scope. For the three-run serving comparisons, request p99 is the median of the p99 values from the three runs.

\paragraph{Recommendation serving.}
All systems use the same physical NVIDIA H100 80GB HBM3 and the same trace of 1,000 fixed, scorable users from the quality evaluation, selected independently of labels.  The main FIR service has a KV budget of 262,144 tokens. The full-text teacher and Ours use FP16 rankers and FP32 output heads; HLLM uses a BF16/FP32 configuration. For the execution-method comparison in Table~\ref{exp:runtime}, we fix the model, item table, hardware, and request trace.

At each concurrency level, we report median throughput across three closed-loop replays, each lasting at least 30 seconds and completing the full trace. Quality is measured using each user's first response in the first run. On this trace, the full-text teacher on vLLM and Ours achieve NDCG@10 of 0.5801 and 0.5833, respectively.

\paragraph{Rating serving.}
HLLM-rating and Ours are compared on the same GPU, a different NVIDIA H100 80GB HBM3 from the recommendation experiments. HLLM-rating uses FP32. Ours uses an FP16 ranker and vector table, an FP32 five-star head, and four CPU input compilation processes. Each run contains 100 requests, and we report the median across three runs.

For the full development-set quality evaluation of Ours ($m=1$) on rating prediction, the ranker and output head use FP32; vectors stored offline in FP16 are converted to FP32 after loading.

\paragraph{VID comparison.}
Table~\ref{exp:vid} fixes FIR ($m=8$), an FP16 ranker, an FP32 output head, and one CPU compilation process. At each concurrency level, a single replay contains 70 requests and 1,900 candidates, with a mixture of $N\in\{5,20,50,100\}$ and a total KV budget of 131,072 tokens. The item table occupies 2,171,764,736 bytes per instance. The maximum probability difference across the four input paths is 0.000547, below the preset threshold of 0.001.

\section{Additional Results for Rating Prediction}
\label{app:rating}
\setcounter{table}{0}
\renewcommand{\thetable}{\thesection\arabic{table}}
\renewcommand{\theHtable}{\thesection.\arabic{table}}

This section examines how rating predictions rank items within each user and how additional history affects these predictions. We also test the effect of changing historical feedback and compare long and short input suffixes.

\subsection{Within-User Ranking and History Length}
\label{app:within-user}
\paragraph{Within-user ranking protocol.}
We rank each user's 5--10 target items by their predicted expected ratings to assess whether the rating model correctly orders items according to the user's observed preferences.
This diagnostic differs from the next-item recommendation evaluation in Table~\ref{exp:recommendation-quality}: all target items here have observed star ratings.

NDCG@3~\citep{jarvelin2002cumulated} uses gain $2^{y-1}-1$, where $y$ is the observed rating, with prediction ties evaluated using the average gain of the tied group~\citep{mcsherry2008tied}.
Pairwise agreement considers target pairs with different observed ratings and assigns 0.5 to prediction ties.
Top-1 regret is the difference between the highest observed rating and the average observed rating of items tied for first in the predicted ranking.
NDCG@3 and pairwise agreement use 758 users with differing target ratings, while Top-1 regret uses all 1,000 users.

\paragraph{Overall ranking results.}
DuoLM improves within-user ranking over History prior across all three metrics (Table~\ref{app:tab:within-user}).
The teacher model achieves the highest NDCG@3 and pairwise agreement.

\appColumnTable{
\centering
\small
\setlength{\tabcolsep}{2pt}
\renewcommand{\arraystretch}{1.10}
\caption{Within-user ranking, averaged equally across history lengths $K\in\{1,3,5,10\}$. Agreement denotes pairwise agreement; regret denotes Top-1 regret. Bold/underlined: best/second-best.}
\label{app:tab:within-user}
\begin{tabular*}{\linewidth}{@{\extracolsep{\fill}}lrrr@{}}
\toprule[0.9pt]
\textbf{Model} & \shortstack{\textbf{NDCG@3}\\$\uparrow$} & \shortstack{\textbf{Agreement}\\$\uparrow$} & \shortstack{\textbf{Regret}\\$\downarrow$} \\
\midrule[0.45pt]
History prior & 0.7207 & 0.5000 & 0.5657 \\
HLLM-rating & \underline{0.7682} & 0.5626 & \textbf{0.4263} \\
Teacher model & \textbf{0.7719} & \textbf{0.5675} & \underline{0.4425} \\
\addlinespace[3pt]
FIE ($m=8$) & \underline{0.7682} & \underline{0.5645} & 0.4588 \\
FIR ($m=8$) & 0.7634 & 0.5535 & 0.4900 \\
DuoLM & 0.7634 & 0.5524 & 0.4640 \\
\bottomrule[0.9pt]
\end{tabular*}
}

\paragraph{Rating prediction across history lengths.}
DuoLM's NLL, both MAE measures, RMSE, and RPS decrease consistently as history length increases (Table~\ref{app:tab:history-length}).
From $K=1$ to $K=10$, expected-rating MAE decreases from 0.6888 to 0.6027, and RMSE decreases from 0.8197 to 0.7334.

\appColumnTable{
\centering
\small
\setlength{\tabcolsep}{2pt}
\renewcommand{\arraystretch}{1.10}
\caption{Results for DuoLM at different history lengths. MAE$_{\mathrm{med}}$ uses the predictive median; MAE$_{\mathrm{exp}}$ and RMSE use the expected rating. NDCG@3 and agreement evaluate within-user ranking. Bold/underlined: best/second-best.}
\label{app:tab:history-length}
\begin{tabular*}{\linewidth}{@{\extracolsep{\fill}}lrrrr@{}}
\toprule[0.9pt]
\textbf{Metric} & $\boldsymbol{K=1}$ & $\boldsymbol{K=3}$ & $\boldsymbol{K=5}$ & $\boldsymbol{K=10}$ \\
\midrule[0.45pt]
\textbf{NLL} $\downarrow$ & 0.9848 & 0.9337 & \underline{0.9061} & \textbf{0.8806} \\
\textbf{MAE}$_{\mathrm{med}}$ $\downarrow$ & 0.5705 & 0.5394 & \underline{0.5294} & \textbf{0.5056} \\
\textbf{MAE}$_{\mathrm{exp}}$ $\downarrow$ & 0.6888 & 0.6473 & \underline{0.6273} & \textbf{0.6027} \\
\textbf{RMSE} $\downarrow$ & 0.8197 & 0.7798 & \underline{0.7569} & \textbf{0.7334} \\
\textbf{RPS} $\downarrow$ & 0.1044 & 0.0975 & \underline{0.0943} & \textbf{0.0911} \\
\midrule[0.45pt]
\textbf{NDCG@3} $\uparrow$ & \textbf{0.7641} & \underline{0.7633} & 0.7631 & 0.7629 \\
\textbf{Agreement} $\uparrow$ & \underline{0.5529} & 0.5521 & \textbf{0.5530} & 0.5516 \\
\bottomrule[0.9pt]
\end{tabular*}
}

\subsection{Feedback Intervention}
\label{app:feedback-intervention}
We test whether predictions depend on historical feedback by fixing the weights of the full-text teacher, FIE ($m=8$), and FIR ($m=8$), and either shuffling the correspondence between history items and ratings or removing the ratings at inference time.
Shuffling preserves item order and the multiset of ratings; removal preserves the history items.
The proportions of inputs actually changed by shuffling are 0\%, 57.6\%, 69.2\%, and 79.0\% for $K=1,3,5,10$, respectively.

\appColumnTable{
\centering
\small
\setlength{\tabcolsep}{2pt}
\renewcommand{\arraystretch}{1.10}
\caption{Feedback intervention. $\Delta$NLL is the intervention result minus the normal-input result, averaged equally across the four values of $K$. Brackets give 95\% user-paired bootstrap intervals from 10,000 resamples. For normal-input NLL, bold and underlined values are best and second-best. Bold intervention differences have intervals that exclude zero.}
\label{app:tab:feedback-intervention}
\begin{tabular*}{\linewidth}{@{\extracolsep{\fill}}lrc@{}}
\toprule[0.9pt]
\textbf{Intervention} & $\boldsymbol{\Delta}$\textbf{NLL} & \textbf{95\% CI} \\
\midrule[0.45pt]
\multicolumn{3}{l}{Full-text teacher\quad NLL $\downarrow$: \textbf{0.9233}} \\
Shuffle & $\mathbf{+0.0042}$ & $[+0.00219, +0.00632]$ \\
Remove & $\mathbf{+0.1051}$ & $[+0.09250, +0.11822]$ \\
\midrule[0.45pt]
\multicolumn{3}{l}{FIE ($m=8$)\quad NLL $\downarrow$: \underline{0.9236}} \\
Shuffle & $\mathbf{+0.0016}$ & $[+0.00028, +0.00293]$ \\
Remove & $\mathbf{+0.1593}$ & $[+0.14059, +0.17877]$ \\
\midrule[0.45pt]
\multicolumn{3}{l}{FIR ($m=8$)\quad NLL $\downarrow$: 0.9249} \\
Shuffle & $\mathbf{+0.0028}$ & $[+0.00125, +0.00432]$ \\
Remove & $\mathbf{+0.1073}$ & $[+0.09381, +0.12128]$ \\
\bottomrule[0.9pt]
\end{tabular*}
}

Shuffling or removing historical ratings increases NLL for the full-text teacher, FIE ($m=8$), and FIR ($m=8$), with removal having a larger effect (Table~\ref{app:tab:feedback-intervention}).
These results show that all three models use historical feedback in their predictions.

\subsection{Input Suffix Templates}
\label{app:tail-template}
We compare the original long input suffix with the short suffix for FIR ($m=1$), keeping the same weights and changing only the input template (Table~\ref{app:tab:tail-template}).

\appColumnTable{
\centering
\small
\setlength{\tabcolsep}{4pt}
\renewcommand{\arraystretch}{1.10}
\caption{Long and short input-suffix templates for FIR ($m=1$). Bold/underlined: best/second-best.}
\label{app:tab:tail-template}
\begin{tabular*}{\linewidth}{@{\extracolsep{\fill}}lrr@{}}
\toprule[0.9pt]
\textbf{Metric} $\downarrow$ & \shortstack{\textbf{Original}\\\textbf{long suffix}} & \shortstack{\textbf{Short suffix}\\\textbf{(main text)}} \\
\midrule[0.45pt]
NLL & \underline{0.9268} & \textbf{0.9263} \\
MAE$_{\mathrm{med}}$ & \textbf{0.5355} & \underline{0.5362} \\
MAE$_{\mathrm{exp}}$ & \underline{0.6430} & \textbf{0.6415} \\
RMSE & \underline{0.7749} & \textbf{0.7725} \\
RPS & \underline{0.0969} & \textbf{0.0968} \\
\bottomrule[0.9pt]
\end{tabular*}
}

We use the short suffix in the main text to reduce the number of input tokens.
\FloatBarrier
\section{Training and Ablation Details}
\label{app:training}
\setcounter{table}{0}
\renewcommand{\thetable}{\thesection\arabic{table}}
\renewcommand{\theHtable}{\thesection.\arabic{table}}

This section details the training settings for the Books ablations and reports additional metrics, paired comparisons, and training costs.

\subsection{Experimental Settings}

\paragraph{Shared settings.}
The readout and loss ablations share a separately trained default model. The seven Books experiments use the same 237,558 training queries, an effective batch size of 32, a microbatch size of 8, and 7,424 unique updates. Modules shared across configurations are initialized from the same weights. Checkpoints are selected on a fixed development subset of 2,000 users, and results are reported on all 5,000 users.

\paragraph{Readouts and continuous input interface.}
Every readout produces one item vector. LastToken uses the Encoder's final-layer hidden state at EOS~\citep{zhang2025qwen3embedding}. MEM1 appends one learnable 1,024-dimensional input vector, while Perceiver1~\citep{jaegle2021perceiver} uses one learnable query and a single layer of eight-head attention to aggregate content.
The continuous input interface RMS-normalizes each readout vector~\citep{zhang2019root} and calibrates its scale:
\[
\begin{gathered}
z=\rho e^{\gamma}\frac{a}{\sqrt{d^{-1}\sum_{j=1}^{d}a_j^2+10^{-6}}},\\
d=1024.
\end{gathered}
\]
Here, $a$ is the readout vector, and $\gamma$ is a learnable scalar log-gain initialized to zero. For recommendation, we compute the reference RMS jointly over the ranker $B$'s native content embeddings for the first 256 distinct training items and fix $\rho=0.0280669$.

\paragraph{Training objectives.}
The default recommendation objective combines recommendation cross-entropy with teacher-to-student distillation~\citep{hinton2015distillingknowledgeneuralnetwork}:
\[
\mathcal{L}=(1-\beta)\mathcal{L}_{\mathrm{CE}}
+\beta\tau^2D_{\mathrm{KL}}(p_T^\tau\|p_S^\tau).
\]
This objective uses the frozen full-text teacher's distribution. Here, $\beta$ is the scheduled distillation weight and $\tau$ is the distillation temperature; table entries for $\beta$ denote its final value.

\paragraph{Books distillation schedule.}
Following the schedule in Appendix~\ref{app:distillation-schedule}, the default Books FIR ($m=1$) recipe uses
$\beta_{\mathrm{final}}=0.5$, $\tau=1$, and
$r_{\mathrm{KD}}=0.1$.
Here, $U$ is the planned number of optimizer updates, $R$ is the ramp length, and $r_{\mathrm{KD}}$ is the fraction of planned updates allocated to the ramp.
One epoch over 237,558 training requests with an effective
batch size of 32 requests gives
\begin{equation}
U=\left\lceil\frac{237{,}558}{32}\right\rceil=7{,}424,
\qquad R=743.
\label{eq:books-distillation-schedule}
\end{equation}
The distillation weight increases linearly over the first 743 optimizer updates: it starts at
$\beta_1\approx6.73\times10^{-4}$, reaches $0.5$ at
update 743, and remains at $0.5$ thereafter.

\paragraph{Self-alignment.}
The self-alignment variant extends the default recommendation objective with an output KL loss between two branches of the same $B$ (full-text branch $\|$ compressed branch, at temperature 1) and a hidden-state cosine loss~\citep{sanh2019distilbert}. Gradients propagate through both branches. The output alignment weight decreases from 0.5 to 0.1, the hidden-state alignment weight decreases from 0.1 to 0.02, and the additional full-text CE weight is zero.

\vspace{4pt}
\subsection{Ablation Results and Training Cost}

We compare the selected and final checkpoints separately to examine whether the ablation differences persist at the end of training.

\paragraph{Time accounting.}
Training time covers the update phase. The $\tau=2$ run includes 678 updates replayed after recovery. Self-alignment uses gradient checkpointing~\citep{chen2016training} for $B$, while CE-only retains the teacher-cache reading path.

\appFixedTable{
\centering
\small
\setlength{\tabcolsep}{1pt}
\renewcommand{\arraystretch}{1.10}
\caption{Additional ablation metrics and training update time. Metrics are reported separately for the selected and final checkpoints. Update time is reported once per recipe. Bold/underlined: best/second-best. }
\label{app:tab:training-metrics}
\begin{tabular*}{\linewidth}{@{}>{\raggedright\arraybackslash}p{0.34\linewidth}@{\extracolsep{\fill}}rrrr@{}}
\toprule[0.9pt]
\multicolumn{5}{@{}l}{\textbf{Selected checkpoint}} \\
\midrule[0.45pt]
\textbf{Recipe} & \shortstack{\textbf{HR@1}\\$\uparrow$} & \shortstack{\textbf{HR@10}\\$\uparrow$} & \shortstack{\textbf{MRR@10}\\$\uparrow$} & \shortstack{\textbf{Update}\\\textbf{time (h)}} \\
\midrule[0.45pt]
LastToken/\allowbreak{}default & \textbf{0.4024} & \underline{0.7452} & \textbf{0.5049} & 7.39 \\
MEM1 & 0.3900 & 0.7374 & 0.4982 & 7.54 \\
Perceiver1 & 0.3588 & 0.7320 & 0.4714 & 8.96 \\
\addlinespace[4pt]
CE-only & 0.3748 & \textbf{0.7474} & 0.4884 & 7.37 \\
$\beta=0.9$ & 0.3774 & 0.7284 & 0.4846 & 7.40 \\
$\tau=2$ & \underline{0.3904} & 0.7424 & \underline{0.4985} & 8.06 \\
\addlinespace[4pt]
$+$ Self-alignment & 0.3900 & 0.7414 & 0.4971 & 16.95 \\
\bottomrule[0.9pt]
\end{tabular*}
\par\medskip
\begin{tabular*}{\linewidth}{@{}>{\raggedright\arraybackslash}p{0.34\linewidth}@{\extracolsep{\fill}}rrr@{}}
\toprule[0.9pt]
\multicolumn{4}{@{}l}{\textbf{Final checkpoint}} \\
\midrule[0.45pt]
\textbf{Recipe} & \shortstack{\textbf{HR@1}\\$\uparrow$} & \shortstack{\textbf{HR@10}\\$\uparrow$} & \shortstack{\textbf{MRR@10}\\$\uparrow$} \\
\midrule[0.45pt]
LastToken/\allowbreak{}default & \textbf{0.3922} & 0.7372 & \textbf{0.4984} \\
MEM1 & \underline{0.3904} & 0.7386 & \underline{0.4974} \\
Perceiver1 & 0.3474 & 0.7302 & 0.4627 \\
\addlinespace[4pt]
CE-only & 0.3772 & \textbf{0.7438} & 0.4882 \\
$\beta=0.9$ & 0.3742 & 0.7258 & 0.4801 \\
$\tau=2$ & 0.3856 & 0.7366 & 0.4940 \\
\addlinespace[4pt]
$+$ Self-alignment & 0.3870 & \underline{0.7392} & 0.4943 \\
\bottomrule[0.9pt]
\end{tabular*}
}

At both checkpoints, the default LastToken+CE+KD recipe has higher HR@1, MRR@10, and NDCG@10 than CE-only, while CE-only has slightly higher HR@10 (Tables~\ref{app:tab:training-metrics} and~\ref{app:tab:bootstrap}).

\paragraph{Paired Ablation Comparisons.}
For each ablation, Table~\ref{app:tab:bootstrap} reports NDCG@10 differences (variant minus default) and 95\% confidence intervals from 10,000 user-paired bootstrap~\citep{efron1979bootstrap} resamples, separately for the selected and final checkpoints.

\appColumnTable{
\centering
\small
\setlength{\tabcolsep}{1pt}
\renewcommand{\arraystretch}{1.10}
\caption{NDCG@10 differences relative to the default LastToken+CE+KD recipe and 95\% user-paired bootstrap intervals. Bold differences have intervals excluding zero.}
\label{app:tab:bootstrap}
\begin{tabular*}{\linewidth}{@{}>{\raggedright\arraybackslash}p{0.34\linewidth}@{\extracolsep{\fill}}rl@{}}
\toprule[0.9pt]
\multicolumn{3}{@{}l}{\textbf{Selected checkpoint}} \\
\midrule[0.45pt]
\textbf{Recipe} & $\Delta$\textbf{NDCG@10} & \textbf{95\% CI} \\
\midrule[0.45pt]
MEM1 & $\mathbf{-0.0066}$ & $[-0.0108, -0.0024]$ \\
Perceiver1 & $\mathbf{-0.0286}$ & $[-0.0343, -0.0229]$ \\
\addlinespace[4pt]
CE-only & $\mathbf{-0.0119}$ & $[-0.0174, -0.0065]$ \\
$\beta=0.9$ & $\mathbf{-0.0193}$ & $[-0.0243, -0.0144]$ \\
$\tau=2$ & $\mathbf{-0.0053}$ & $[-0.0099, -0.0007]$ \\
\addlinespace[4pt]
$+$ Self-alignment & $\mathbf{-0.0067}$ & $[-0.0106, -0.0028]$ \\
\midrule[0.45pt]
\multicolumn{3}{@{}l}{\textbf{Final checkpoint}} \\
\midrule[0.45pt]
\textbf{Recipe} & $\Delta$\textbf{NDCG@10} & \textbf{95\% CI} \\
\midrule[0.45pt]
MEM1 & $-0.0004$ & $[-0.0046, +0.0038]$ \\
Perceiver1 & $\mathbf{-0.0290}$ & $[-0.0346, -0.0233]$ \\
\addlinespace[4pt]
CE-only & $\mathbf{-0.0064}$ & $[-0.0115, -0.0012]$ \\
$\beta=0.9$ & $\mathbf{-0.0168}$ & $[-0.0216, -0.0120]$ \\
$\tau=2$ & $-0.0035$ & $[-0.0079, +0.0008]$ \\
\addlinespace[4pt]
$+$ Self-alignment & $-0.0027$ & $[-0.0068, +0.0013]$ \\
\bottomrule[0.9pt]
\end{tabular*}
}

At the selected checkpoint, all six intervals lie below zero. At the final checkpoint, the intervals remain below zero for CE-only, $\beta=0.9$, and Perceiver1. The gaps for MEM1, $\tau=2$, and self-alignment narrow, and their final-checkpoint intervals cross zero.

\paragraph{Training cost.}
Relative to the default recipe, self-alignment increases update time from 7.39 to 16.95 hours without improving NDCG@10. Perceiver1 takes 8.96 hours and also has lower NDCG@10 than the default.

\section{Additional Serving Results}
\label{app:serving}
\setcounter{table}{0}
\renewcommand{\thetable}{\thesection\arabic{table}}
\renewcommand{\theHtable}{\thesection.\arabic{table}}

This section provides detailed serving measurements, covering execution ablations, request latency, the rating comparison with the teacher model, and CPU input preparation.

\subsection{Execution and Input-Path Diagnostics}
\label{app:runtime-diagnostics}

Table~\ref{exp:runtime} reports the exact throughput values underlying Figure~\ref{fig:execution-ablation}, in candidates/s. The execution ablation fixes the recommendation model and request trace, with 100 candidates per request and cross-request prefix reuse disabled.

\appColumnTable{
\centering
\caption{Numerical results for Figure~\ref{fig:execution-ablation}.
Execution ablation with a fixed recommendation model.
Each request has 100 candidates.
Throughput is in candidates/s (higher is better).
$C$ denotes concurrent recommendation requests.
Green/yellow mark the best/second-best execution methods at each concurrency.}
\label{exp:runtime}
\centering
\begingroup
\fontsize{10pt}{12pt}\selectfont
\setlength{\tabcolsep}{3pt}
\renewcommand{\arraystretch}{1.16}

\begin{tabular}{rrrr}
\toprule
\multicolumn{4}{c}{Throughput (candidates/s) $\uparrow$} \\
\addlinespace[2pt]
$C$
& \shortstack{Independent\\scoring}
& MIS
& \shortstack{MIS +\\CUDA Graph} \\
\midrule
1
& 520.53
& \cellcolor{tableOneSecond}2,503.55
& \cellcolor{tableOneBest}8,870.36 \\
8
& 921.45
& \cellcolor{tableOneSecond}9,480.19
& \cellcolor{tableOneBest}20,740.56 \\
32
& 881.61
& \cellcolor{tableOneSecond}19,883.01
& \cellcolor{tableOneBest}25,506.11 \\
64
& 851.48
& \cellcolor{tableOneSecond}24,953.95
& \cellcolor{tableOneBest}26,403.98 \\
\bottomrule
\end{tabular}

\endgroup
}

Table~\ref{exp:vid} reports a separate rating input-path diagnostic with FIR ($m=8$) and mixed $N\in\{5,20,50,100\}$. This diagnostic uses 10 users and one replay per concurrency setting (Appendix~\ref{app:serving-protocol}). Request sizes are in bytes and throughput is in candidates/s. FP32/FP16 denote the precision of transferred vectors, and ``cache'' denotes cross-request prefix caching. Figure~\ref{fig:vid-input-path} plots these values in kB ($1\,\mathrm{kB}=1{,}000$ bytes) and thousands of candidates/s, with rounded labels.

\appColumnTable{
\centering
\small
\setlength{\tabcolsep}{2pt}
\renewcommand{\arraystretch}{1.12}
\caption{VID input paths for FIR ($m=8$), with mixed
$N\in\{5,20,50,100\}$.
Bytes denotes mean request size; throughput is in candidates/s.
FP32/FP16 denote transferred-vector precision.
VID+cache adds cross-request prefix caching.
Green/yellow mark best/second-best values, including ties.
Full values for Figure~\ref{fig:vid-input-path}.}
\label{exp:vid}
\begin{tabular*}{\linewidth}{@{\extracolsep{\fill}}lrrrr@{}}
\toprule
Metric & FP32 & FP16 & VID & \shortstack{VID+\\cache} \\
\midrule
Bytes $\downarrow$
& 1,227,361
& \cellcolor{tableOneSecond}618,812
& \cellcolor{tableOneBest}11,842
& \cellcolor{tableOneBest}11,842 \\
\midrule
$C=1$ $\uparrow$
& 711.60
& 731.94
& \cellcolor{tableOneBest}1,058.68
& \cellcolor{tableOneSecond}1,021.97 \\
$C=8$ $\uparrow$
& 1,744.17
& 1,659.23
& \cellcolor{tableOneBest}2,903.34
& \cellcolor{tableOneSecond}2,784.68 \\
\bottomrule
\end{tabular*}
}

In this diagnostic, FP16 transfer reduces request size relative to FP32 but has lower throughput at $C=8$. VID+cache also has lower measured throughput than VID at both reported concurrency levels.

\subsection{Request Latency}
\label{app:latency}

MIS lowers the p99 latency of recommendation requests, and CUDA Graph further reduces latency at low concurrency (Figure~\ref{app:fig:recommendation-latency} and Table~\ref{app:tab:recommendation-latency}). At $C=1$, p99 falls from 485.01 ms with independent execution to 43.43 ms with MIS and 12.35 ms with MIS+CUDA Graph.

\appColumnTable{
\centering
\small
\setlength{\tabcolsep}{1pt}
\renewcommand{\arraystretch}{1.10}
\renewcommand{\thetable}{\thesection\arabic{table}a}
\renewcommand{\theHtable}{\thesection.\arabic{table}a}
\caption{Request p99 latency (ms) for Books recommendation, corresponding to Tables~\ref{exp:recommendation-throughput} and~\ref{exp:runtime}. Lower is better. Bold/underlined: best/second-best.}
\label{app:tab:recommendation-latency}
\begin{tabular*}{\linewidth}{@{}>{\raggedright\arraybackslash}p{0.24\linewidth}@{\extracolsep{\fill}}rrrr@{}}
\toprule[0.9pt]
\textbf{System} & $\boldsymbol{C=1}$ & $\boldsymbol{C=8}$ & $\boldsymbol{C=32}$ & $\boldsymbol{C=64}$ \\
\midrule[0.45pt]
\shortstack[l]{Full-text\\teacher on\\vLLM} & 354.87 & 2,099.20 & 8,155.82 & 17,093.82 \\
\addlinespace[3pt]
\shortstack[l]{Full-text\\teacher on\\SGLang} & 706.46 & 4,138.95 & 14,701.82 & 28,189.04 \\
\addlinespace[3pt]
HLLM & \underline{23.77} & \underline{46.32} & \textbf{55.89} & \textbf{75.54} \\
\addlinespace[4pt]
\shortstack[l]{Ours:\\independent\\execution} & 485.01 & 1,236.90 & 4,024.63 & 8,104.30 \\
\addlinespace[3pt]
Ours: MIS & 43.43 & 268.19 & 424.85 & 539.88 \\
\addlinespace[3pt]
\shortstack[l]{Ours:\\MIS+CUDA\\Graph} & \textbf{12.35} & \textbf{43.89} & \underline{389.27} & \underline{536.90} \\
\bottomrule[0.9pt]
\end{tabular*}
}

\appColumnFigure{
\centering
\includegraphics[width=\linewidth]{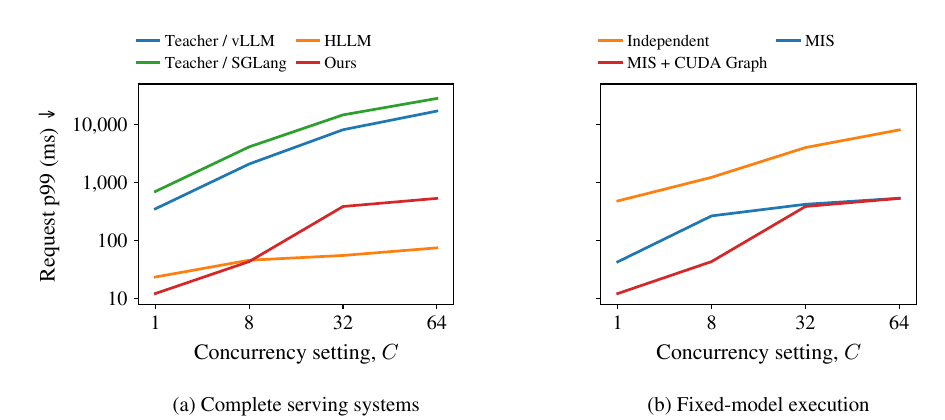}
\caption{Request p99 latency for Books recommendation. Left: complete serving systems. Right: execution methods for the same model. Each request scores 100 candidates. vLLM and SGLang denote full-text teacher deployments; Ours denotes MIS with CUDA Graph. Both latency axes use a logarithmic scale. Exact values appear in Table~\ref{app:tab:recommendation-latency}.}
\label{app:fig:recommendation-latency}
}

The separate rating comparison uses Ours ($m=1$) and HLLM-rating, with the serving setup described in Appendix~\ref{app:serving-protocol}.

\appColumnTable{
\centering
\small
\setlength{\tabcolsep}{3pt}
\renewcommand{\arraystretch}{1.10}
\addtocounter{table}{-1}
\renewcommand{\thetable}{\thesection\arabic{table}b}
\renewcommand{\theHtable}{\thesection.\arabic{table}b}
\caption{Request p99 latency (ms $\downarrow$) for Books rating prediction. Bold/underlined: best/second-best.}
\label{app:tab:rating-latency}
\begin{tabular*}{\linewidth}{@{\extracolsep{\fill}}crr@{}}
\toprule[0.9pt]
$\boldsymbol{C}$ & \textbf{HLLM-rating} & \textbf{Ours ($m=1$)} \\
\midrule[0.45pt]
1  & \textbf{25.14} & \underline{84.44} \\
8  & \textbf{48.28} & \underline{136.09} \\
16 & \textbf{53.68} & \underline{207.20} \\
32 & \textbf{66.57} & \underline{330.95} \\
\bottomrule[0.9pt]
\end{tabular*}
}

\subsection{Rating Serving Comparison with the Full-Text Teacher}
\label{app:teacher-rating-serving}

We compare FIR ($m=8$) with the full-text teacher on SGLang using a mixed $N\in\{5,20,50,100\}$ rating workload. At each concurrency level, we perform three replays of 700 requests and 19,000 candidates each. Both systems run on the same H100 with shared input preparation, MIS, FP16 rankers, and FP32 output heads, with cross-request prefix caching disabled. FIR uses offline vectors and a persistent tokenization directory; the full-text teacher uses a bounded tokenization cache reset before each run. Both implementations pass numerical checks with a maximum probability difference threshold of 0.001.

On this mixed-$N$ workload, FIR achieves $2.36\times$ the full-text teacher's throughput at $C=1$ and $4.52\times$ at $C=8$, measured in requests/s (Table~\ref{app:tab:teacher-rating-throughput}). For requests with $N=100$, p50 latency falls from 417.42 ms with the full-text teacher to 92.03 ms with FIR at $C=1$, and from 2,427.76 to 322.11 ms at $C=8$.

\appColumnTable{
\centering
\small
\setlength{\tabcolsep}{3pt}
\renewcommand{\arraystretch}{1.10}
\caption{Rating-serving throughput (requests/s $\uparrow$). Bold/underlined: best/second-best.}
\label{app:tab:teacher-rating-throughput}
\begin{tabular*}{\linewidth}{@{\extracolsep{\fill}}lrr@{}}
\toprule[0.9pt]
\textbf{Model} & $\boldsymbol{C=1}$ & $\boldsymbol{C=8}$ \\
\midrule[0.45pt]
\shortstack[l]{Full-text teacher\\on SGLang} & \underline{7.5954} & \underline{11.3411} \\
FIR ($m=8$) & \textbf{17.8960} & \textbf{51.3040} \\
\midrule[0.45pt]
\shortstack[l]{FIR/full-text teacher\\throughput ratio} & $2.36\times$ & $4.52\times$ \\
\bottomrule[0.9pt]
\end{tabular*}
}

\subsection{CPU Input Preparation}
\label{app:cpu}

\appColumnFigure{
\centering
\includegraphics[width=\linewidth]{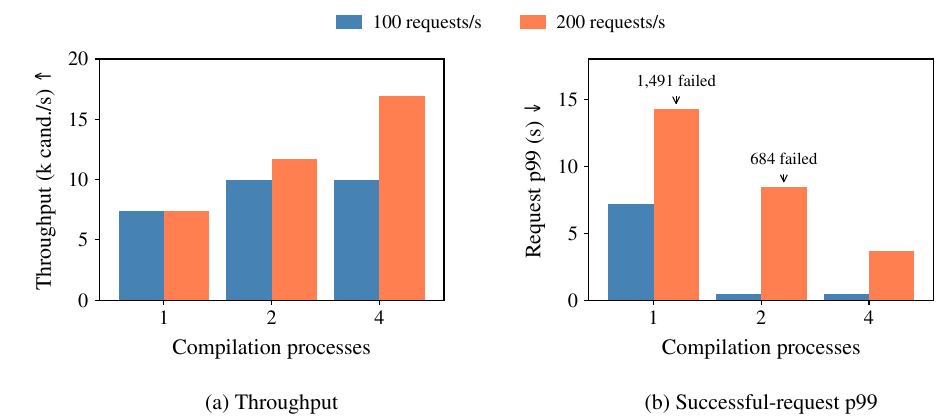}
\caption{Effect of CPU compilation processes at two request rates. Bars show throughput and successful-request p99 latency. Table~\ref{app:tab:cpu} reports backlog growth and failures.}
\label{app:fig:cpu-input}
}

We compare 1, 2, and 4 compilation processes under open-loop arrivals of 100 and 200 requests/s, with the rating model fixed at $m=1$, $K=10$, $N=100$, and a concurrency limit of 32. Each measurement uses a 20-second arrival window, followed by a drain phase to complete the remaining requests. We cycle through 100 request samples and recompile the inputs each time.

Throughput includes the drain phase, and p99 covers successful requests only. Backlog growth is the mean outstanding request count in the last quarter minus that in the second quarter of the arrival window.

\appColumnTable{
\centering
\small
\setlength{\tabcolsep}{1pt}
\renewcommand{\arraystretch}{1.10}
\caption{Open-loop input processing. Procs: compilation processes. Bold/underlined: best/second-best within each arrival rate. Backlog growth and failures are not ranked.}
\label{app:tab:cpu}
\begin{tabular*}{\linewidth}{@{}>{\raggedright\arraybackslash}p{0.39\linewidth}@{\extracolsep{\fill}}rrr@{}}
\toprule[0.9pt]
\multicolumn{4}{@{}l@{}}{\textbf{Rate: 100 requests/s}} \\
\addlinespace[2pt]
& \multicolumn{3}{c@{}}{\textbf{Procs}} \\
\textbf{Metric} & \textbf{1} & \textbf{2} & \textbf{4} \\
\midrule[0.45pt]
\shortstack[l]{Throughput $\uparrow$\\(candidates/s)} & 7,358.81 & \underline{9,962.70} & \textbf{9,971.89} \\
\addlinespace[3pt]
p99 $\downarrow$ (ms) & 7,193.63 & \underline{488.25} & \textbf{474.90} \\
\addlinespace[3pt]
\shortstack[l]{Backlog growth\\(requests)} & 246.994 & 0.868 & 1.280 \\
\addlinespace[3pt]
Failures & 0 & 0 & 0 \\
\midrule[0.45pt]
\multicolumn{4}{@{}l@{}}{\textbf{Rate: 200 requests/s}} \\
\addlinespace[2pt]
& \multicolumn{3}{c@{}}{\textbf{Procs}} \\
\textbf{Metric} & \textbf{1} & \textbf{2} & \textbf{4} \\
\midrule[0.45pt]
\shortstack[l]{Throughput $\uparrow$\\(candidates/s)} & 7,402.85 & \underline{11,683.26} & \textbf{16,969.58} \\
\addlinespace[3pt]
p99 $\downarrow$ (ms) & 14,274.90 & \underline{8,437.53} & \textbf{3,691.89} \\
\addlinespace[3pt]
\shortstack[l]{Backlog growth\\(requests)} & 123.034 & 252.563 & 302.942 \\
\addlinespace[3pt]
Failures & 1,491 & 684 & 0 \\
\bottomrule[0.9pt]
\end{tabular*}
}

At 100 requests/s, two processes reach approximately 10,000 candidates/s, with little further throughput gain from four (Figure~\ref{app:fig:cpu-input} and Table~\ref{app:tab:cpu}). All three configurations complete without failures; backlog growth is 246.994 requests with one process, compared with 0.868 and 1.280 with two and four.

At 200 requests/s, four processes raise throughput to 16,969.58 candidates/s and complete without failures, while backlog grows by 302.942 requests. One and two processes record 1,491 and 684 failures, respectively. Thus, the higher throughput at this arrival rate accompanies continued backlog growth.

\FloatBarrier

\section{Supplementary Method Details}
\label{app:method-details}

This section provides further details on teacher training, student optimization, and the MIS implementation.

\subsection{Expert Teacher Training}

\label{app:teacher-training}

Let $\widetilde X_{q,j}$ denote the teacher's input embedding
sequence for candidate $j$ in request $q$, constructed using
the text template in Section~\ref{sec:learning}.
With parameters $\phi=(\theta_T,w_T)$, the teacher produces
\begin{equation}
t_{q,j}(\phi)
=
w_T^{\mathsf T}
\big[T_{\theta_T}(\widetilde X_{q,j})\big]_{\mathrm{last}},
\label{eq:teacher-score}
\end{equation}
where $\mathrm{last}$ denotes the final token of the scoring suffix.

For each request, $y_q$ indexes the observed target among
$K_q$ candidates. We train the teacher with cross-entropy:
\begin{equation}
\mathcal L_T(\phi)
=
-\frac{1}{|\mathcal B_T|}
\sum_{q\in\mathcal B_T}
\log
\frac{\exp(t_{q,y_q}(\phi))}
{\sum_{j=1}^{K_q}\exp(t_{q,j}(\phi))},
\label{eq:teacher-loss}
\end{equation}
where $\mathcal B_T$ is a batch of training requests.
After training, we freeze $\phi$ and cache the raw scores
for the exact histories and ordered candidate groups
used in student training.

\subsection{Student Training}
\label{app:reproducibility}

\paragraph{Distillation Schedule.}
\label{app:distillation-schedule}

The weight $\beta_u$ in Eq.~\eqref{eq:student-objective} controls
the balance between label supervision and distillation from the
cached teacher scores. Let $u$ count optimizer updates after
gradient accumulation, and let $U$ be the planned number of such updates.
We linearly increase the distillation weight to
$\beta_{\mathrm{final}}\in[0,1]$ over the first $R$ updates
and then keep it fixed:
\begin{equation}
\beta_u
=
\beta_{\mathrm{final}}
\min\left(1,\frac{u}{R}\right),
\qquad u=1,\ldots,U.
\label{eq:weight-schedule}
\end{equation}
The implementation computes the ramp length as
\begin{equation}
R=\max\bigl(1,\lceil U r_{\mathrm{KD}}\rceil\bigr),
\label{eq:distillation-ramp-length}
\end{equation}
where $r_{\mathrm{KD}}\in[0,1]$ is the fraction of planned
updates allocated to the ramp, giving $1\leq R\leq U$.
When $r_{\mathrm{KD}}=0$, we have $R=1$ and apply
$\beta_{\mathrm{final}}$ from the first update.
For $R>1$, the first update uses
$\beta_1=\beta_{\mathrm{final}}/R$.
Resumed runs continue the schedule from the saved update
index plus one.

\paragraph{Student Score Gradients.}
\label{app:score-gradients}

For request $q$, differentiating the per-request objective
$\ell_u(q)$ in Eq.~\eqref{eq:student-objective} with respect to
candidate score $s_{q,j}$ gives a label residual and a teacher residual:
\begin{equation}
\begin{aligned}
\frac{\partial\ell_u(q)}{\partial s_{q,j}}
={}&(1-\beta_u)\bigl(p_{\mathrm S,q,j}^{(1)}
-\mathbf 1[j=y_q]\bigr)\\[4pt]
&+\beta_u\tau\bigl(p_{\mathrm S,q,j}^{(\tau)}
-p_{\mathrm T,q,j}^{(\tau)}\bigr),
\end{aligned}
\label{eq:score-gradient}
\end{equation}
where $\mathbf 1[\cdot]$ is the indicator function.
The factor $\tau$ in the teacher residual comes from the
$\tau^2$ loss scaling and the $1/\tau$ factor from differentiating
the temperature-scaled softmax.
The label residual favors the observed target, whereas the teacher
residual matches the teacher's candidate-group distribution.
Both supervision signals propagate through the ranker and
continuous input interface to the encoder; the teacher and reference
scale $\rho$ remain fixed.

\subsection{Multi-Item Scoring}
\label{app:mis-extensions}

We implement multi-item scoring (MIS) on top of SGLang~\citep{NEURIPS2024_724be447} to score multiple candidates in a single forward pass. Each request packs a shared history prefix followed by candidate segments separated by transport delimiters. An item-aware causal attention mask allows each candidate to attend to the shared prefix and its own segment while blocking cross-candidate attention. The prefix KV states are computed once per request and shared across candidates. To execute these packed requests, we pass the prefix length and candidate-boundary metadata to FlashInfer's prefill kernels~\citep{MLSYS2025_dbf02b21}.

We assign token positions within each candidate as if it followed the shared prefix directly, independent of the other candidates. For our recommendation model, the scalar head reads the final candidate hidden state immediately before each trailing delimiter. We exclude the endpoint of the shared history prefix and return one score per candidate in the original order.

\paragraph{Candidate Attention.}
\label{app:mis-attention}

Within this packed implementation, the recommendation model
represents each candidate with one continuous vector
(Section~\ref{sec:scoring}). The attention formula below computes
this candidate row's output using the prefix rows and the candidate
row itself, following the visibility rule in Eq.~\eqref{eq:candidate-visibility}
and the logical positions defined in Section~\ref{sec:serving}.
For one request and attention head at a given layer, let $K_P,V_P$
denote prefix key/value matrices and $q_j,k_j,v_j$ the candidate's
query, key, and value rows.
Queries and keys incorporate positional transformations at the
assigned logical positions.
With head width $d_h$, candidate attention~\citep{NIPS2017_3f5ee243} is
\begin{equation}
o_j=\operatorname{softmax}\!\left(
\frac{q_j(K_P\oplus k_j)^{\mathsf T}}{\sqrt{d_h}}
\right)(V_P\oplus v_j).
\label{eq:shared-attention}
\end{equation}
Here, $\oplus$ concatenates rows along the sequence dimension.

\end{document}